\documentclass{amsart}

\newcommand {\supplus}{\mathop{{\supset}\llap{\raise 
0.5pt\hbox{\normalfont\small+}\hskip 0.5pt}}} 

\newcommand {\subplus}{\mathop{{\subset}\llap{\raise 
0.5pt\hbox{\normalfont\small+}\hskip 0.5pt}}}  

\newcommand {\Cee}    {{\mathbb  C}}

\newcommand {\Zee}    {{\mathbb  Z}}

\newcommand {\fa}     {{\mathfrak{a}}}

\newcommand {\fab}    {{\mathfrak{ab}}} 
\newcommand {\fag}    {{\mathfrak{ag}}}

\newcommand {\faut}   {{\mathfrak{aut}}} 
\newcommand {\fb}     {{\mathfrak{b}}}
\newcommand {\fc}    {{\mathfrak{c}}}

\newcommand {\fder}   {{\mathfrak{der}}}   %

\newcommand {\fd}     {{\mathfrak{d}}}

\newcommand {\fg}     {{\mathfrak{g}}}    %
\newcommand {\fgl}    {{\mathfrak{gl}}}  %
\newcommand {\fh}     {{\mathfrak{h}}}

\newcommand {\fk}     {{\mathfrak{k}}}

\newcommand {\fle}    {{\mathfrak{le}}}
\newcommand {\fn}     {{\mathfrak{n}}}
\newcommand {\fns}     {{\mathfrak{ns}}}
\newcommand {\fo}     {{\mathfrak{o}}}
\newcommand {\fosp}   {{\mathfrak{osp}}}
\newcommand {\fp}    {{\mathfrak{p}}}   %
\newcommand {\fpe}    {{\mathfrak{pe}}}   %

\newcommand {\fpo}    {{\mathfrak{po}}}

\newcommand {\fpsq}   {{\mathfrak{psq}}}
\newcommand {\fq}     {{\mathfrak{q}}}     
\newcommand {\fr}     {{\mathfrak{r}}}

\newcommand {\fs}     {{\mathfrak{s}}}

\newcommand {\fsb}    {{\mathfrak{sb}}}

\newcommand {\fsh}    {{\mathfrak{sh}}}

\newcommand {\fsl}    {{\mathfrak{sl}}}
\newcommand {\fsle}   {{\mathfrak{sle}}}

\newcommand {\fsp}    {{\mathfrak{sp}}}
\newcommand {\fspe}   {{\mathfrak{spe}}}
\newcommand {\fspo}   {{\mathfrak{spo}}}

\newcommand {\fsvect} {{\mathfrak{svect}}}

\newcommand {\fvect}  {{\mathfrak{vect}}}   %
\newcommand {\fvir}   {{\mathfrak{vir}}}

\newcommand {\fz}     {{\mathfrak{z}}}

\newcommand {\cal} {\mathcal}

\newcommand {\cL}     {{\cal L}}

\def \opname#1#2%
  {\expandafter\newcommand \csname #1\endcsname {{\mathop{#2}\nolimits}}}

\newcommand{\rmname}[1]
  {\expandafter\newcommand \csname #1\endcsname {{\operatorname{#1}}}}

\newcommand{\rmnameii}[2]
  {\expandafter\newcommand \csname #1\endcsname {{\operatorname{#2}}}}

\rmname{act}
\rmname{Ad}
\rmname{Add}
\rmname{ad}
\rmname{Alt}
\rmname{alt}
\rmname{Ann}
\rmname{antidiag}
\rmname{Ber}
\rmname{ber}
\rmname{Br}
\rmname{card}
\rmname{ch}
\rmname{Char}
\rmname{cem}
\rmname{cj}
\rmname{Cliff}
\rmname{cntr}
\rmname{codim}
\rmname{coind}
\rmname{const}
\rmname{col}
\rmname{cork}
\rmname{cpr}
\rmname{diag}
\rmnameii{Div}{div}
\rmname{Def}
\rmname{Der}
\rmname{Dim}
\rmname{End}
\rmname{Even}
\rmname{Ext}
\rmname{gr}
\rmname{Hom}
\rmname{HT}
\rmnameii{Ht}{ht}
\rmname{hwt}
\rmname{Id}
\rmname{id}
\rmname{ind}
\rmname{Ind}
\rmname{Inf}
\rmname{irr}
\rmname{Le}
\rmname{Lie}
\rmname{lwt}
\rmname{mult}
\rmname{Mor}
\rmname{nm}
\rmname{Ob}
\rmname{Odd}
\rmname{Osc}
\rmname{per}
\rmname{Pic}
\rmname{pr}
\rmname{pro}
\rmname{Prime}
\rmname{Proj}
\rmname{prt}
\rmname{pt}
\rmname{Q}
\rmname{qet}
\rmname{qtr}
\rmname{rd}
\rmname{rk}
\rmname{row}
\rmname{Res}
\rmname{salt}
\rmname{Sch}
\rmname{SBr}
\rmname{scalar}
\rmname{Ser}
\rmname{sign}
\rmname{Smbl}
\rmname{spin}
\rmname{ssym}
\rmname{str}
\rmname{st}
\rmname{sgn}
\rmname{sq}
\rmname{symm}
\rmname{supp}
\rmname{Supp}
\rmname{St}
\rmname{Spec}
\rmname{Spm}
\rmname{tr}
\rmname{vpt}
\rmname{weyl}
\rmname{Weyl}
\rmname{Witt}

\opname{vvol}  {{v\hspace{-0.1ex}o\hspace{-0.02ex}l\/}}
\opname{pnt}  {\text{\normalfont pt}}
\opname{Span} {{Span}}
\opname{slim} {\overline{\lim}}
\opname{Vol}  {{V\hspace{-0.55ex}o\hspace{-0.02ex}l\/}}
\opname{QVol} {{Q\hspace{-0.3ex}V\hspace{-0.55ex}o\hspace{-0.02ex}l\/}}
\opname{PoVol}{{P\hspace{-0.35ex}o\hspace{-0.25ex}V\hspace{-0.55ex}o\hspace{-0.02ex}l\/}}
\opname{BVol} {{B\hspace{-0.2ex}V\hspace{-0.55ex}o\hspace{-0.02ex}l\/}}
\opname{Par}  {{P\hspace{-0.3ex}a\hspace{-0.05ex}r\/}}

\rmname{Mat}
\rmname{Bil}
\rmname{Diff}
\rmname{Ker}
\rmname{Herm}
\rmname{Coker}
\rmname{Conn}
\rmname{Covect}
\rmname{Vect}
\rmname{Int}

\rmnameii {IM} {Im}
\rmnameii {RE} {Re}

\opname{Aut} {{A\hspace{-0.2ex}u\hspace{-0.1ex}t\/}}
\opname{GL} {{G\hspace{-0.3ex}L}}
\opname{SL} {{S\hspace{-0.3ex}L}}
\opname{Exp} {{E\hspace{-0.2ex}x\hspace{-0.1ex}p\/}}
\opname{GQ} {{G\hspace{-0.2ex}Q}}
\opname{OSp} {{O\hspace{-0.25ex}S\hspace{-0.15ex}p\/}}
\opname{Out} {{O\hspace{-0.25ex}u\hspace{-0.15ex}t\/}}
\opname{Spp} {{S\hspace{-0.2ex}p\/}}
\opname{SpO} {{S\hspace{-0.2ex}p\hspace{-0.02ex}O\/}}
\opname{Pe} {{P\hspace{-0.25ex}e\/}}
\opname{SPe} {{S\hspace{-0.25ex}P\hspace{-0.25ex}e\/}}
\opname{Spin} {{S\hspace{-0.25ex}p\hspace{-0.05ex}i\hspace{-0.1ex}n\/}}
\opname{Iso} {{I\hspace{-0.25ex}s\hspace{-0.1ex}o\/}}
\opname{SSPe} {{S\hspace{-0.25ex}S\hspace{-0.15ex}P\hspace{-0.25ex}e\/}}
\opname{PeU} {{P\hspace{-0.25ex}e\hspace{-0.1ex}U\/}}
\opname{QU} {{Q\hspace{-0.15ex}U\/}}
\opname{U} {{U\/}}

\opname{cGQ} {{\cal G \hspace{-0.2em} Q \/}}
\opname{cSL} {{\cal S \hspace{-0.2em} L \/}}
\opname{cGL} {{\cal G \hspace{-0.2em} L \/}}
\opname{cOSp} {{\cal O \hspace{-0.2em} S \hspace{-0.3em} \it p\/}}
\opname{cPe} {{\cal P \hspace{-1.5pt} \it e\/}}
\opname{cVect} {{\cal V \hspace{-1.5pt} \it e\hspace{-0.1ex}c\hspace{-0.1ex}t\/}}
\opname{cVol} {{\cal V \hspace{-1.5pt} \it o\hspace{-0.1ex}l\/}}
\opname{cAut} {{\cal A \hspace{-0.2em} \it u\hspace{-0.1em}t\/}}
\opname{cCovect} {{\cal C \hspace{-1.5pt}
     \it o\hspace{-0.1ex}v\hspace{-0.1ex}e\hspace{-0.1ex}c\hspace{-0.1ex}t\/}}
\opname{CW} {{C\hspace{-0.15ex}W}}

\opname {Ab}   {{\sf Ab}}
\opname {Alg}   {{\sf Alg}}
\opname {ASch}  {{\sf Aff\;Sch}}
\opname {Funct}   {{\sf Funct}}
\opname {Gr}   {{\sf Gr}}
\opname {Grf}  {{{\sf Gr}_f}}
\opname {Mods}   {{\sf mods}}
\opname {Rings}   {{\sf Rings}}
\opname {Salg}   {{\sf Salg}}
\opname {Sets} {{\sf Sets}}
\opname {SSMan} {{\sf SMan}}
\opname {Top}  {{\sf Top}}
\opname {Vebun}   {{\sf Vebun}}

\newcommand {\ev} {{\bar0}}
\newcommand {\od} {{\bar1}}
\newcommand {\eps} {\varepsilon}
\newcommand {\degree}  {{}^\circ}
\newcommand {\tto} {\longrightarrow}
\newcommand {\pder}[1] {{\frac{\partial}{\partial {#1}}}}
\newcommand {\pderf}[2] {{\frac{\partial {#1}}{\partial {#2}}}}

\newcommand {\bcdot}   {\mathbin{\hbox{\raise.4ex\hbox{\bf.}}}} 

\newcommand {\secno} {}
\newcommand {\ssecfont} {\normalfont\bf}

\newtheorem{Theorem}{\secno Theorem}

\newtheorem{Proposition}[Theorem]{\secno Proposition}

\newtheorem{Statement}[Theorem]{\secno Statement}
\newtheorem{Problem}[Theorem]{\secno Problem}

\newenvironment {th*}[1]
    {\gdef\thname{#1} \begin{thn}}%
    {\end{thn}}
\newtheorem{thn}[Theorem] {\thname}

\theoremstyle{definition}

\newtheorem{Convention}[Theorem]{\secno Convention}

\newenvironment {ex*}[1]
    {\gdef\thname{#1} \begin{exn}}%
    {\end{exn}}
\newtheorem{exn}[Theorem]{\thname}

\theoremstyle{remark}

\newenvironment {rem*}[1]
    {\gdef\thname{#1} \begin{remn}}%
    {\end{remn}}
\newtheorem{remn}[Theorem]{\thname}

\newcommand {\ssec}{\subsection*}

\newcommand {\ssbegin}[2]
  {\def \secno {\gdef \secno {}{\ssecfont #1. }}%
   \begin{#2}}
\begin{document}
\title[Defining Relations for Lie superalgebras]{Defining Relations for Lie
Superalgebras\\ with Cartan matrix}

\author{Pavel Grozman, Dimitry Leites} 

\address{Dept. of Math., Univ. of Stockholm, Roslagsv. 101, 
Kr\"aftriket hus 6, 
S-106 91, Stockholm, Sweden\\ e-mail:mleites\@matematik.su.se}

\thanks{Financial support of the Swedish Institute and
NFR are gratefully acknowledged.}

\begin{abstract} The notion of defining relations is well-defined for a nilpotent
Lie (super)algebra. One of the ways to present a simple Lie algebra is,
therefore, by splitting it into the direct sum of a maximal diagonalizing
(commutative) subalgeba and 2 nilpotent subalgebras (positive and negative). The
relations obtained for finite dimensional Lie algebras are neat; they are called
{\it Serre relations} and can be encoded via an integer symmetrizable matrix, the {\it
Cartan matrix}, which, in turn, can be encoded by means of a graph, the {\it Dynkin
diagram}. The complete set of relations for Lie algebras with an arbitrary Cartan
matrix is unknown.

We completely describe presentations of Lie
superalgebras with Cartan matrix if they are simple
$\Zee$-graded of polynomial growth. Such matrices can be neither integer
nor symmetrizable. There are {\it non-Serre} relations encountered. In certain cases
there are infinitely many relations.

Our results are applicable to the Lie algebras with the same Cartan
matrices as the Lie superalgebras considered.
\end{abstract}



\maketitle

\section*{Introduction}
This paper is the direct continuation of [K2], [LSS], [L1], [LSe]. In [LSe] the the case of the
simplest (for computations) base is considered and non-Serre relations are first written. Though
we are studying the Lie superalgebras with Cartan matrix, we give examples of Lie
superalgebras of the other types, to illustrate the geometry related with some of our algebras.

An explicit presentation of simple Lie superalgebras became urgently needed in connection
with $q$-quantization of Lie superalgebras: straightforward generalization of Drinfeld's
results (who used Serre relations) is insufficient here, cf. [FLV]. After [FLV], there
appeared a paper [Sch] $q$-quantizing $\fsl(m|n)$ and, to an extent, $\fosp(m|2n)$. Though
long, the paper [Sch] lacks explicit formulas for $\fosp(m|2n)$ and even for $\fsl(m|n)$ the
explicit form of the general formulas is not given. Nowhere, so far, are the defining 
relations for all systems of simple roots of the exceptional Lie superalgebras written
down ({\it some} systems are considered in [Y]). 

Here for all simple Lie superlagebras $\fg(A)$ with Cartan matrix $A$ we list the defining relations
for each system of simple roots. If $\dim\fg(A)<\infty$ and for for
$\Zee$-graded Lie superlagebras $\fg(A)$ of polynomial growth with a symmetrizable $A$ this list is
complete (proof is the same as in [K1]); in the other cases its completeness is conjectured. Cartan
matrices can be neither integer nor symmetrizable. There are {\it non-Serre} relations encountered.
In certain cases there are infinitely many relations. We will consider the
$q$-quantized versions elsewhere. 

When Kac's method of the proof ([K1]) failed, and to derive our conjectures, we used the package
[G]. To describe the defining relations for $\fg(A)$ with symmetrizable matrices more general than
those considered by Gaber and Kac (see [K1]) or nonsymmetrizable matrices distinct from those
considered here is an open problem.

\section*{\S 0. Background}

\ssec{0.0. Linear algebra in superspaces. Generalities}
Superization has certain subtleties, often disregarded or expressed too briefly, cf. [L].
We will dwell on them a bit.

A {\it superspace} is a $\Zee /2$-graded space; for a superspace
$V=V_{\ev}\oplus V_{\od }$ denote by $\Pi (V)$ another copy of the same
superspace: with the shifted parity, i.e., $(\Pi(V))_{\bar i}= V_{\bar i+\od}$. The
{\it superdimension} of $V$ is $\dim V=p+q\varepsilon$, where
$\varepsilon^2=1$ and $p=\dim V_{\ev}$,
$q=\dim V_{\od }$. (Usually, $\dim V$ is expressed as a pair $(p,q)$ or
$p|q$; this obscures the fact that $\dim V\otimes W=\dim V\cdot \dim W$.)

A superspace structure in $V$ induces the superspace structure in the space $\End
(V)$. A {\it superalgebra} is a superspace $A$ with an even multiplication map
$m:A\otimes A\longrightarrow A$.

A {\it basis of a superspace} is by definition a basis consisting of {\it
homogeneous} vectors; let
$\Par=(p_1, \dots, p_{\dim V})$ be an ordered collection of their parities.
We call $\Par$ the {\it format} of $V$. A square {\it supermatrix} of format (size)
$\Par$ is a $\dim V\times \dim V$ matrix whose $i$th row and $i$th column are of
the same parity $p_i$. The matrix unit $E_{ij}$ is supposed to be of parity
$p_i+p_j$ and the bracket of supermatrices (of the same format) is defined via Sign Rule: {\it if something of
parity $p$ moves past something of parity $q$ the sign $(-1)^{pq}$ accrues; the
formulas defined on homogeneous elements are extended to arbitrary ones via
linearity}. For example: setting $[X, Y]=XY-(-1)^{p(X)p(Y)}YX$ we get the notion
of the supercommutator and the ensuing notion of the Lie superalgebra (that
satisfies the superskew-commutativity and super Jacobi identity).

We do not usually use the sign $\wedge$ for differential forms on supermanifolds: in what
follows we assume that the exterior differential is odd and the differential forms
constitute a supercommutative superalgebra; we keep using it on manifolds, sometimes,
not to diviate too far from the conventional notations. 

Usually, $\Par$ is of the form $(\ev , \dots, \ev , \od , \dots, \od )$. Such a
format is called {\it standard}. The nonstandard formats are vital in the classification of
systems of simple roots; the corresponding defining relations are distinct.

\ssec{0.1.} The {\it general linear} Lie superalgebra of all supermatrices of size $\Par$ is
denoted by $\fgl(\Par)$, usually, $\fgl(\ev, \dots, \ev, \od, \dots, \od)$ is
abbreviated to $\fgl(\dim V_{\bar 0}|\dim V_{\bar 1})$. Any matrix from
$\fgl(m|n)$ (i.e., in the standard format) can be expressed as the sum of its even
and odd parts: 
$$
\begin{pmatrix}A&B\\ C&D\end{pmatrix}=\begin{pmatrix}A&0\\
0&D\end{pmatrix}+\begin{pmatrix}0&B\\ C&0\end{pmatrix},\quad\text{where}\quad
p\left(\begin{pmatrix}A&0\\ 0&D\end{pmatrix}\right)=\ev, \; p\left(\begin{pmatrix}0&B\\
C&0\end{pmatrix}\right)=\od.
$$

More generally, we can consider matrices with the elements from a
(usually, supercommutative) superalgebra $C$. Then the parity of the matrix with only one
nonzero $i, j$-th entry $X_{i, j}\in C$is equal to $p(i)+p(j)+p(X_{i, j})$.

The {\it supertrace} is the map $\fgl
(\Par)\longrightarrow
\Cee$,
$(A_{ij})\mapsto \sum (-1)^{p_{i}}A_{ii}$. The superspace of supertraceless matrices
constitutes the {\it special linear} Lie subsuperalgebra $\fsl(\Par)$.

{\bf Superalgebras that preserve bilinear forms: two types}. To the linear map
$F:V\longrightarrow W$ of superspaces there corresponds the dual map
$F^*:W^*\longrightarrow V^*$ of the dual superspaces; if $A$ is the supermatrix
corresponding to $F$ in a format
$\Par$, then to $F^*$ the {\it supertransposed} matrix $A^{st}$ corresponds:
$$
(A^{st})_{ij}=(-1)^{(p_{i}+p_{j})(p_{i}+p(A))}A_{ji}.
$$

The supermatrices $X\in\fgl(\Par)$ such that 
$$
X^{st}B+(-1)^{p(X)p(B)}BX=0\quad \text{for a homogeneous matrix $B\in\fgl(\Par)$}
$$
constitute the Lie superalgebra $\faut (B)$ that preserves the bilinear form on $V$ with
matrix $B$. If $B$ corresponds to a nondegenerate supersymmetric form, then for the
canonical form of its matrix in the standard format we can select either $B_{ev}$ or
$B'_{ev}$ (each canonical form has its (dis)advantages):
$$
B_{ev}(m|2n)= \begin{pmatrix} 
1_m&0\\
0&J_{2n}
\end{pmatrix},\quad \text{where $J_{2n}=\begin{pmatrix}0&1_n\\-1_n&0\end{pmatrix}$,
or}\; B'_{ev}(m|2n)= \begin{pmatrix} 
\antidiag (1, \dots , 1)&0\\
0&J_{2n}
\end{pmatrix}. 
$$
The usual notation for $\faut (B_{ev}(m|2n))$ is $\fosp^{sy}(m|2n)$.
(Observe that the passage from $V$ to $\Pi (V)$ sends the supersymmetric forms to
superskew-symmetric ones, preserved by the \lq\lq
symplectico-orthogonal" Lie superalgebra $\fosp^{sk}(m|2n)$ which
is isomorphic to $\fosp^{sy}(m|2n)$ but has a different matrix realization. We never use
notation $\fsp'\fo (2n|m)$ in order not to confuse with the special Poisson
superalgebra. To understand this statement, recall that on the superspace of bilinear
forms $Bil(V)$ there is an involution $u: B\mapsto B^u$ which on matrices acts as
follows:
$$
B=\begin{pmatrix} 
B_{11}&B_{12}\\
B_{21}&B_{22}
\end{pmatrix}\mapsto B^u=\begin{pmatrix} 
B_{11}^t&B_{21}\\
B_{12}&B_{22}^t
\end{pmatrix}. 
$$

In the standard format the matrix realizations of these algebras
are: 
$$
\begin{matrix} 
\fosp^{sy}(m|2n)=\left\{\left (\begin{matrix} E&Y&-X^t\\
X&A&B\\
Y^t&C&-A^t\end{matrix} \right)\right\};\quad \fosp^{sk}(m|2n)=
\left\{\left(\begin{matrix} A&B&X\\
C&-A^t&Y^t\\
Y^t&-X^t&E\end{matrix} \right)\right\}, \\
\text{where}\; 
\left(\begin{matrix} A&B\\
C&-A^t\end{matrix} \right)\in \fsp(2n),\qquad E\in\fo(m)\;
\text{and}\; {}^t \; \text{is the usual transposition}.\end{matrix} 
$$

A nondegenerate supersymmetric odd bilinear form $B_{odd}(n|n)$ can be
reduced to a canonical form whose matrix in the standard format is 
$J_{2n}$. A canonical form of the superskew odd nondegenerate form in the
standard format is $\Pi_{2n}=\begin{pmatrix} 0&1_n\\1_n&0\end{pmatrix}$.
The usual notation for $\faut (B_{odd}(\Par))$ is $\fpe(\Par)$. The passage
from $V$ to $\Pi (V)$ sends the supersymmetric forms to superskew-symmetric
ones and establishes an isomorphism $\fpe^{sy}(\Par)\cong\fpe^{sk}(\Par)$. This Lie
superalgebra is called, as A.~Weil suggested, {\it periplectic}, i.e., odd-plectic. The
matrix realizations in the standard format of these superalgebras is:
$$
\begin{matrix}
\fpe ^{sy}\ (n)=\left\{\begin{pmatrix} A&B\\
C&-A^t\end{pmatrix}, \; \text{where}\; B=-B^t,\; 
C=C^t\right\};\\
\fpe^{sk}(n)=\left\{\begin{pmatrix}A&B\\ C&-A^t\end{pmatrix}, \;
\text{where}\; B=B^t, \; C=-C^t\right\}.
\end{matrix}
$$

The {\it special periplectic} superalgebra is $\fspe(n)=\{X\in\fpe(n): \str
X=0\}$.

\ssec{0.2. Vectoral Lie superalgebras. The standard realization} The elements of the
Lie algebra $\cL=\fder\;\Cee [[u]]$ are considered as vector fields. The Lie algebra $\cL$ has
only one maximal subalgebra
$\cL_0$ of finite codimension (consisting of the fields that vanish at the origin). The
subalgebra $\cL_0$ determines a filtration of $\cL$: set
$$
\cL_{-1}=\cL;\quad \cL_i =\{D\in \cL_{i-1}: [D, \cL]\subset\cL_{i-1}\}\; \text{for
}i\geq 1.
$$
The associated graded Lie algebra $L=\mathop{\oplus}\limits_{i\geq -1}L_i$, where
$L_i=\cL_{i}/\cL_{i+1}$, consists of the vector fields with {\it polynomial} coefficients.

Unlike Lie algebras, simple vectoral {\it super}algebras possess {\it several} maximal
subalgebras of finite codimension. 

{\bf 1) General algebras}. Let $x=(u_1, \dots , u_n, \theta_1, \dots ,
\theta_m)$ the $u_i$ are even indeterminates and the $\theta_j$ are odd ones.
The Lie superalgebra $\fvect (n|m)$ is $\fder\; \Cee[x]$; it is called {\it the
general vectoral superalgebra}. \index{$\fvect$ general vectoral Lie
superalgebra}\index{ Lie superalgebra general vectoral}

\begin{rem*}{Remark} Sometimes we write $\fvect (x)$ or even $\fvect (V)$ if
$V=\Span(x)$ and use similar notations for the subalgebras of $\fvect$ introduced
below. Algebraists sometimes abbreviate $\fvect (n)$ and $\fs\fvect (n)$ to $W_n$
(in honor of Witt) and $S_n$, respectively.
\end{rem*}

{\bf 2) Special algebras}. The {\it divergence}\index{divergence} of the field
$D=\sum\limits_if_i\pder{u_{i}} + \sum\limits_j
g_j\pder{\theta_{j}}$ is the function (in our case: a
polynomial, or a series) 
$$
\Div D=\sum\limits_i\pderf{f_{i}}{u_{i}}+
\sum\limits_j (-1)^{p(g_{j})}
\pderf{g_{i}}{\theta_{j}}.
$$

$\bullet$ The Lie superalgerba $\fsvect (n|m)=\{D \in \fvect (n|m): \Div D=0\}$
is called the {\it special} or {\it divergence-free vectoral superalgebra}.
\index{$\fsvect$ general vectoral Lie superalgebra}\index{ Lie superalgebra
special vectoral}\index{ Lie superalgebra
divergence-free}

It is clear that it is also possible to describe $\fsvect$ as $\{ D\in \fvect
(n|m): L_D\vvol _x=0\}$, where $\vvol_x$ is the volume form with constant
coefficients in coordinates $x$ and $L_D$ the Lie derivative with respect to
$D$. 

$\bullet$ The Lie superalgerba $\fsvect_{\lambda}(0|m)=\{D \in \fvect (0|m):
\Div (1+\lambda\theta_1\cdot \dots \cdot
\theta_m)D=0\}$ --- the deform of $\fsvect(0|m)$ --- is called the
{\it special} or {\it divergence-free vectoral superalgebra}. It is cleat that 
$\fsvect_{\lambda}(0|m)\cong \fsvect_{\mu}(0|m)$ for
$\lambda\mu\neq 0$. Observe that $p(\lambda)\equiv m\mod 2$, i.e., for odd $m$
the parameter of deformation $\lambda$ is odd.

{\bf 3) The algebras that preserve Pfaff
equations and differential 2-forms}. Set 
$$
\tilde \alpha = dt +\sum\limits_{1\leq i\leq n}(p_idq_i - q_idp_i)\ +
\sum\limits_{1\leq j\leq m}\theta_jd\theta_j\quad\text{and}\quad 
\tilde \omega=d\tilde \alpha.
$$
(Here we set $u=(t, p_1, \dots , p_n, q_1, \dots , q_n)$.) The form $\alpha_1$ is called {\it
contact}, the form $\omega_0$ is called {\it symplectic}.\index{form differential
contact}\index{form differential symplectic} Sometimes it is more convenient to redenote
the $\theta$'s and set 
$$
\xi_j=\frac{1}{\sqrt{2}}(\theta_{j}-i\theta_{r+j});\quad \eta_j=\frac{1}{
\sqrt{2}}(\theta_{j}+i\theta_{r+j})\; \text{ for}\; j\leq r= [m/2]\; (\text{here}\;
i^2=-1),\quad
\theta =\theta_{2r+1} 
$$ 
and in place of $\tilde \omega$ or $\tilde \alpha$ take $\alpha$
and $\omega=d\alpha$, respectively, where 
$$
\begin{array}{rcl}
\alpha=&dt+\sum\limits_{1\leq i\leq n}(p_idq_i-q_idp_i)+
\sum\limits_{1\leq j\leq r}(\xi_jd\eta_j+\eta_jd\xi_j)&
\text{ if }\ m=2r\\
\alpha=&dt+\sum\limits_{1\leq i\leq n}(p_idq_i-q_idp_i)+
\sum\limits_{1\leq j\leq r}(\xi_jd\eta_j+\eta_jd\xi_j) +\theta d\theta&\text{ if
}\ m=2r+1.\end{array} 
$$

The Lie superalgebra that preserves the {\it Pfaff equation}
\index{Pfaff equation} $\alpha=0$, i.e., the superalgebra
$$
\fk (2n+1|m)=\{ D\in \fvect (2n+1|m): L_D\alpha=f_D\alpha\}, 
$$
(here $f_D\in \Cee [t, p, q, \xi]$ is a polynomial determined by $D$) is
called the {\it contact superalgebra}.\index{$\fk$ contact superalgebra}
\index{Lie superalgebra contact} The Lie superalgebra
$$
\begin{array}{c}
\fpo (2n|m)=\{ D\in \fk (2n+1|m): L_D\alpha_1=0\}\end{array}
$$
is called the {\it Poisson} superalgebra.\index{$\fpo$ Poisson superalgebra}
(A geometric interpretation of the
Poisson superalgebra: it is the Lie superalgebra that preserves the connection
with form $\alpha$ in the line bundle over a symplectic supermanifold
with the symplectic form $d\alpha$.) 

\ssec{0.3. Generating functions} A laconic way to describe
$\fk$ and its subalgebras is via generating functions.

Odd form $\alpha$. For $f\in\Cee [t, p, q, \xi]$ set\index{$K_f$
contact vector field} \index{$H_f$ Hamiltonian vector field}: 
$$
K_f=\triangle(f)\pder{t}-H_f +
\pderf{f}{t} E, 
$$
where
$E=\sum\limits_i y_i
\pder{y_{i}}$ (here the $y$ are all the
coordinates except $t$) is the {\it Euler operator} (which counts the degree
with respect to the $y$), $\triangle (f)=2f-E(f)$, and $H_f$ is the
hamiltonian field with Hamiltonian $f$ that preserves $d\alpha_1$: 
$$
H_f=\sum\limits_{i\leq n}(\pderf{f}{p_i}
\pder{q_i}-\pderf{f}{q_i}
\pder{p_i}) -(-1)^{p(f)}\left(\sum\limits_{j\leq m}\pderf{
f}{\theta_j} \pder{\theta_j}\right ) , \; \; f\in \Cee [p,
q, \theta]. 
$$

The choice of the form $\alpha$ instead of $\alpha$ only affects the
form of $H_f$ that we give for $m=2k+1$:
$$
H_f=\sum\limits_{i\leq n} (\pderf{f}{p_i}
\pder{q_i}-\pderf{f}{q_i}
\pder{p_i}) -(-1)^{p(f)}\sum\limits_{j\leq
k}(\pderf{f}{\xi_j} \pder{\eta_j}+
\pderf{f}{\eta_j} \pder{\xi_j}+
\pderf{f}{\theta} \pder{\theta}), \;
\; f\in \Cee [p, q, \xi, \eta, \theta]. 
$$

Since 
$$
L_{K_f}(\alpha)=K_1(f)\cdot\alpha_1,\eqno{(0.1)}
$$
it follows that $K_f\in \fk (2n+1|m)$. 

$\bullet$ To the supercommutator $[K_f, K_g]$ there
correspond {\it contact bracket}\index{Poisson bracket}\index{contact bracket}
of the generating functions:
$$
[K_f, K_g]=K_{\{f, g\}_{k.b.}}.
$$
The explicit formulas for the contact brackets are as follows. Let us first
define the brackets on functions that do not depend on $t$.

The {\it Poisson bracket} $\{\cdot , \cdot\}_{P.b.}$ (in the realization with the form
$\omega_0$) is given by the formula 
$$
\begin{array}{c}
\{f, g\}_{P.b.}=\sum\limits_{i\leq n}\ (\pderf{f}{p_i}\ 
\pderf{g}{q_i}-\ \pderf{f}{q_i}\ 
\pderf{g}{p_i})-(-1)^{p(f)}\sum\limits_{j\leq m}\ 
\pderf{f}{\theta_j}\ \pderf{g}{\theta_j}\end{array}
$$
and in the realization with the form
$\omega'_0$ for $m=2k+1$ it is given by the formula 
$$
\begin{array}{c}
\{f, g\}_{P.b.}=\sum\limits_{i\leq n}\ (\pderf{f}{p_i}\ 
\pderf{g}{q_i}-\ \pderf{f}{q_i}\ 
\pderf{g}{p_i})-(-1)^{p(f)}[\sum\limits_{j\leq m}( 
\pderf{f}{\xi_j}\ \pderf{
g}{\eta_j}+\pderf{f}{\eta_j}\ \pderf{
g}{\xi_j})+\pderf{f}{\theta}\ \pderf{
g}{\theta}].
\end{array} 
$$

Then
$$
\{ f, g\}_{k.b.}=\triangle (f)\pderf{g}{t}-\pderf{f}
{t}\triangle (g)-\{ f, g\}_{P.b.}.
$$

The Lie superalgebras of {\it Hamiltonian fields}\index{Hamiltonian
vector fields} (or {\it Hamiltonian 
superalgebra}) and its special subalgebra (defined only if $n=0$) are
$$
\fh (2n|m)=\{ D\in \fvect (2n|m):\ L_D\omega_0=0\}\; \text{ and} \;
\fsh (m)=\{ D\in \fh (0|m): \Div D=0\}.
$$

It is not difficult to prove the following isomorphisms (as superspaces): 
$$
\fk (2n+1|m)\cong\Span(K_f: f\in \Cee[t, p, q, \xi]);\quad
\fh (2n|m)\cong\Span(H_f: f\in
\Cee [p, q, \xi]).
$$

\begin{rem*}{Remark} 1) It is obvious that the Lie superalgebras of the
series $\fvect$, $\fsvect$, $\fh$ and $\fpo$ for $n=0$ are finite dimensional.

2) A Lie superalgebra of the series $\fh$ is the quotient of the Lie
superalgebra $\fpo$ modulo the one-dimensional center
$\fz$ generated by constant functions.
Similarly, $\fle$ and $\fsle$ are the quotients of $\fb$ and $\fsb$,
respectively, modulo the one-dimensional (odd) center $\fz$ generated by
constant functions. 

3) There are analogues of the contact and hamiltonian series with an even 1-form, [L].
\end{rem*}

Set $\fspo (m)=\{ K_f\in \fpo (0|m):\int fv_\xi=0\}$; clearly, $\fsh (m)=\fspo
(m)/\fz$.

\ssec{0.4. Nonstandard realizations} In [LSh] we proved that the
following are all the nonstandard gradings of the Lie superalgebras
indicated. Moreover, the gradings in the series
$\fvect$ induce the gradings in the series $\fsvect$, and $\fsvect\degree$; the
gradings in $\fk$ induce the gradings in $\fpo$, $\fh$. In what follows we
consider $\fk (2n+1|m)$ as preserving Pfaff eq. $\alpha =0$.
The standard realizations are marked by $(*)$; note that (bar several exceptions for small $m,
n$) it corresponds to the case of the minimal codimension of ${\cal L}_0$. It corresponds to
$r=0$. There are also several exceptional nonstandard regradings; they are listed in sec. 2.6.

\small
$$
\begin{tabular}{|c|c|}
\hline
Lie superalgebra & its $\Zee$-grading \\ 
\hline
$\fvect (n|m; r)$, & $\deg u_i=\deg \xi_j=1$ for any $i, j$\hskip 5 cm
$(*)$\\ 
\cline{2-2}
$ 0\leq r\leq m$ & $\deg \xi_j=0$ for $1\leq j\leq r;\
\deg u_i=\deg \xi_{r+s}=1$ for any $i, s$ \\ 
\hline
\hline
$\fk (2n+1|m; r)$, $0\leq r\leq [\frac{m}{2}]$ & $\deg t=2$, $\deg p_i=\deg q_i=
\deg \xi_j=\deg \eta_j=\deg \theta_k=1$ for any $i, j, k$\qquad $(*)$ \\ 
\cline{2-2}
& $\deg t=\deg \xi_i=2$, $\deg 
\eta_{i}=0$ for $1\leq i\leq r\leq [\frac{m}{2}]$; \\
&$\deg p_i=\deg q_i=\deg \theta_{j}=1$ for
$j\geq 1$ and all $i$\\ 
\hline
$\fk(1|2m; m)$ & $\deg t =\deg \xi_i=1$, $\deg 
\eta_{i}=0$ for $1\leq i\leq m$ \\ \hline
\end{tabular}
$$

Observe that the Lie superalgebras corresponding to different values of $r$
are isomorphic as abstract Lie superalgebras, but as filtered ones they are
distinct.

\ssec{0.5. Stringy superalgebras}
These superalgebras are particular cases of the Lie algebras of vector fileds, 
namely, those that preserve a structure on a what physicists call superstring, 
i.e., a supermanifold associated with a vector bundle on a circle. These
superalgebras themselves are \lq\lq stringy" indeed: as modules over the Witt
algebra they are direct sums of several modules --- strings.

Let $\varphi$ be an angle parameter on a circle, $t= exp(i \varphi)$. A {\it
stringy superalgebra} is the algebra of derivations of either of the two
supercommutative superalgebras
$$
R^{L}(n)=\Cee [t^{-1}, t, \xi _{1}, \ldots , \xi _{n}]\quad\text{or}\quad R^{M}(n) =\Cee [t^{-1}, t, \xi _{1}, \ldots, \xi _{n-1}, 
\sqrt{t} \xi ].
$$ 

$R^{L}(n)$ is the superalgebra of complex-valued functions expandable into finite
Fourier series or, as superscript indicates, Laurent series. These functions are
considered of the real supermanifold $S^{1|n}$ associated with the
rank $n$ trivial bundle over the circle. We can forget about $\varphi$ and think in
terms of $t$ considered as the even coordinate on $(\Cee^*)^{1|n}$.

$R^{M}(n)$ is the superalgebra of complex-valued functions (expandable into finite
Fourier series) on the supermanifold $S^{1|n-1, M}$ associated with the Whitney
sum of the M\"obius bundle and the rank $n-1$ trivial one. Since, as is well-known
from Differential Geometry, the Whitney sum of two M\" obius bundles is isomorphic to
the trivial bundle of rank 2, it suffices to consider one M\"obius summand.

Introduce analogues of
$\fvect$, $\fsvect$, $\fsvect^{0}$ by substituting
$R^{L}(n)$ instead of $R(n) = \Cee [t, \xi _{1}, \ldots, \xi _{n}]$:
$$
\begin{aligned}
\fvect ^{L}(n) &= \fder~ R^{L}(n);\\
\fsvect^{L}_\lambda(n) &= \{ D\in \fvect ^{L}(n)\ :\Div(t^\lambda D) = 0\}\\
\fk ^{L}(n) &= \{D \in\fvect ^{L}(n): D (\alpha _{1}) = f_ D \alpha _{1}\; 
\text{for}\; \alpha _{1}=dt + \sum \xi _{i} d\xi _{i}\; \text{and}\; f_D \in
R^{L}(n) \}.
\end{aligned} 
$$
 
The same arguments as for $\fk(2m+1|n)$, prove that the elements that constitute
$\fk^{L}(n)$ are generated by functions, and the formula for $K_{f}$ is the
same as for $\fk (1|n)$ with the only difference: $f\in R^{L}(n)$.

\begin{rem*}{Exercise}
The algebras ${\fvect}^{M}(n)$ and ${\fsvect}^{M}_\lambda(n)$ obtained by replacing
$R(n)$ with $R^{M}(n)$ are isomorphic to ${\fvect}^{L}(n)$ and
${\fsvect}^{L}_{\lambda-\frac 12}(n)$, respectively. Moreover,
${\fsvect}^{L}_{\lambda}(n)\cong {\fsvect}^{L}_{\mu}(n)$ if and only if
$\lambda-\mu\in\Zee$. 
\end{rem*}

If $\lambda\in\Zee$, the Lie superalgebra ${\fsvect}^{L}_{\lambda}(n)$ has a
simple ideal of codimension $\eps^n$:
$$
0\tto {\fsvect}\degree{}^{L}_{\lambda}(n)\tto {\fsvect}^{L}_{\lambda}(n)\tto
\xi_1\cdot\dots\cdot\xi_n\partial_t\tto 0.
$$ 

\ssec{0.6. Distinguished stringy superalgebras. Nontrivial central extensions}
Define the {\it residue} on $S^{1|n}$ setting
$$
\Res: \Vol \longrightarrow
\Cee \, , \;\;\; f\Vol_{t, \xi} \mapsto \text{ the coefficient of }\
\frac{\xi_{1} \ldots \xi_{n}}{t} \ \text{ in the expantion of }\ f.
$$
A simple Lie superalgebra is called {\it distinguished} if it has a
nontrivial central extension. The following are all nontrivial central
extensions:

$$ 
\begin{tabular}{|c|c|c|}
\hline
algebra & cocycle & The name of the extended algebra \\
\hline
$\fk^{L}(1|0)$ & $K_{f}, K_{g}\mapsto \Res fK_1^3(g)$ & Virasoro or $\; \fvir$ \\
\hline
$\left. \begin{matrix}\fk^{L}(1|1) \\ \fk^{M}(1) \end{matrix}\right\}$
 & $K_{f}, K_{g}\mapsto \Res fK_{\theta }(K_{1})^2(g)$
 & $\begin{matrix} \text{Neveu-Schwarz or} \; \fns \\
  \text{Ramond or}\; \fr \end{matrix}$ \\
\hline
$ \left.\begin{matrix}\fk^{L}(1|2)\\ \fk^{M}(2) \end{matrix}\right\}$
 & $K_{f}, K_{g}\mapsto \Res fK_{\theta_1 }K_{\theta_2}K_{1}(g)$
 & $\begin{matrix}\text{2-Neveu-Schwarz or}\;\fns(2) \\ \text{2-Ramond or}\;\fr(2)\end{matrix}$ \\
\hline
$\left.\begin{matrix}\fk ^{L}(1|3) \\ \fk^{M}(3)\end{matrix} \right\}$
 & $K_{f}, K_{g}\mapsto \Res fK_{\xi }K_{\theta }K_{\eta }(g)$
 & $\begin{matrix}\text{3-Neveu-Schwarz or}\; \fns(3) \\ \text{3-Ramond or}\; \fr (3)\end{matrix}$ \\
\hline
$\left. \begin{matrix}\fk^{L\circ}(4) \\ \fk^{M}(4)
\end{matrix}\right\}$ & $K_{f}, K_{g}\mapsto \begin{matrix}(1) &
\Res fK_{\theta_{1}}K_{\theta_{2}}K_{\theta_{3}}K_{\theta _{4}}(K_{1})^{-1} (g) \\ 
  (2) & \Res f(tK_{t^{-1}}(g)) \\ 
  (3) & \Res fK_{1}(g)\end{matrix}$
 & $\begin{matrix}(1)\left\{\begin{matrix}\text{4-Neveu-Schwarz}=\fns(4)\\ 
\text{4-Ramond}=\fns(4)\end{matrix}\right.\\ 
  (2)\left\{\begin{matrix}\text{$4'$-Neveu-Schwarz}=\fns(4')\\ 
\text{$4'$-Ramond}=\fns(4')\end{matrix}\right.\\
  (3)\left\{\begin{matrix}\text{$4^0$-Neveu-Schwarz}=\fns(4^0)\\ 
\text{$4^0$-Ramond}=\fns(4^0)\end{matrix}\right.\end{matrix}$ \\ 
\hline 
& the restrictions of the above cocycle (3) & \\
$\fvect ^{L}(1| 2)$ & $D_1=f\pder{t}+g_{1}\pder{ \xi
_{1}}+g_{2} \pder{ \xi _{2}} \;
D_2=\tilde{f}\pder{ t}+\tilde{g}_{1}\pder{ \xi
_{1}}+\tilde{g}_{2}\pder{
\xi_{2}}$& $\widehat{\fvect}^{L}(1|2)$ \\ 
&$\mapsto\;\Res(g_{1}\tilde{g}_{2}'-g_{2}\tilde{g}_{1}'(-1)^{p(D_{1})p(D_{2})})$&\\
\hline
$\fsvect ^{L}_{\lambda }(1| 2)$& the restrictions of the above 
& $\widehat{\fsvect}^{L}_{\lambda}(1|2)$ \\
\hline  \end{tabular}  
$$
To see the formulas of the last two lines better, recall, that explicitely, the embedding
$\fvect(1|k)\tto\fk(1|2k)$ is given by the following formula in which
$\Phi=\sum\xi_i\eta_i$:
$$
\begin{matrix} 
f(\xi)x^n\partial_x\mapsto
(-1)^{p(f)}\frac{1}{2^n}f(\xi)(x+\Phi)^n\\
f(\xi)x^n\partial_i\mapsto
(-1)^{p(f)}\frac{1}{2^n}f(\xi)\eta_i(x+\Phi)^n\\
\end{matrix} 
$$
Recall that $\fsvect^{L}_{\lambda}(1|2)$ is singled out by the formulla 
$$
f\partial_x+\sum f_i\partial_i\in \fsvect^{L}_{\lambda}(1|2)\quad\text{if and only
if }\quad \lambda f=-x\Div D.
$$

\ssec{0.7. Twisted loop superalgebras}
Let $\fg$ be a simple finite dimensional Lie superalgebra, $\varphi$ an
automorphisme of finite order $k$, let $\varepsilon$ be a primitive root of 1 of
degree $k$. The automorphism $\varphi$ determines a $\Zee/k\Zee$-grading on
$\fg$ that we will denote by $\fg = \oplus_{\ev \leq i \leq
\overline{k-1}} ~\fg_{i}$, where
$$ 
\fg _{i} =\fg _{i}(\varphi) =\{ g \in \fg : \;
\varphi(g) = \varepsilon^{i} g \}.
$$
The Lie superalgebra
$$
\fg ^{(1)} = \fg _{\id}^{(1)} = \fg \otimes \Cee [t^{-1}, t];
$$
is called a {\it loop superalgebra}.\index{loop superalgebra} The Lie superalgebra
$$
\fg _{\varphi}^{(k)} =\mathop{\oplus}\limits_{ m \in \Zee , \; \ev \leq j \leq \overline{k-1}}\;\;
\fg_{j}t^{mk+j} 
$$
is called a\index{loop superalgebra twisted} {\it twisted loop
superalgebra}. (The maps of the circle
somewhere are loops. So, the term ``loop algebra" stems from the possibility to
identify $ \fg ^{(1)}$ and $\fg _{\varphi}^{(k)}$ with the Lie superalgebra of
$\fg $-valued functions on the circle expandable into finite Fourier series.) 

In applications we encounter nontrivial central extentions of (twisted)
loop superalgebras rather than the superalgebras themselves. The span of such an
extension and the operator $t\frac{d}{dt}$ will be called a {\it Kac--Moody
superalgebra}.\index{Kac--Moody superalgebra}

\begin{Theorem} {\em (Serganova, see [L], v. 22) a)} For a simple finite dimensional $\fg$ 
and an automorphism $\varphi \in \Out\fg$ the superalgebra $\fg _{\varphi}^{(k)}$
does not contain any nontrivial ideal homogeneous with respect to the
$\Zee$-grading defined by the formulas $\deg g =0$ for $g \in \fg $,
$\deg t=1$.

{\em b)} Let $\varphi_{1}$ and $\varphi_{2}$ be two automorphisms of
$\fg$ of orders $k_{1}$ and $k_{2}$, respectively. If
$\varphi_{1} \varphi_{2}^{-1} \in \Aut^{0} (\fg )$, then
$\fg_{\varphi_{1}}^{(k_{1})} \cong \fg _{\varphi_{2}}^{(k_{2})}$.
\end{Theorem}

\ssec{0.8. Central extensions of (twisted) loop superalgebras} There are two types of central
extensions: one is associated with a nondegenerate supersymmetric invariant bilinear form on
$\fg$ (for the even form this is the straightforward generalization of the Kac--Moody cocycle),
the other one is the series: functions with values in the extensions of $\fg$.
$$
\begin{tabular}{|c|c|c|}
\hline
\text{what is}&$ \text{the cocycle}$&$
\text{The name of the}$\\
\text{extended}&&\text{result of the extention}\\
&&of $\fg^{(m)}_{\varphi }$\\
\hline
&$ 1)~(X, Y)\mapsto \Res ~
B(X, \frac{dY}{dt})$&$ E_{B}(\fg^{(m)}_{\varphi })$, \\
$\fg^{(m)}_{\varphi }$&$\text{ where $B$ is a symmetric bilinear form on }\fg$&\\
&$2) ~c_{i}: X, Y\mapsto \Res ~t^{i}c(X, Y)$&
$ E_{c_{i}}(\fg^{(m)}_{\varphi })$, \\
&$\text{ where $c$ is a nontrivial cocycle on } \; \; \fg $&\\
\hline\end{tabular}
$$

\ssec{0.9. Exceptional algebras and $\fg(A)$} All the Lie superalgebras described in this section are
$\Zee$-graded of polynomial growth, i.e., of the form $\fg=\mathop{\oplus}\limits_{i\in\Zee}\fg_i$,
where $[\fg_i, \fg_j]\subset\fg_{i+j}$ (this means that $\fg$ is graded); $\dim\fg_i<\infty$ for all
$i$ and $\sum\limits_{|i|<n}\dim\fg_i$ grows as a polynomial in $n$. For the, so far limited,
applications of the algebras whose growth is faster than a polynomial one see [K1].

Observe that the exceptional Lie algebras are rather difficult to describe. The same is true for
Lie superalgebras. The only consise way to describe them is with the help of the Cartan matrix,
i.e., to present them as $\fg(A)$.

\section*{\protect \S 1. What is $\fg(A)$ and how to present it}
First, recall, how to construct a Lie algebra from a Cartan matrix. Let $A=(a_{ij})$
be an arbitrary complex $n \times n$ matrix of rank $l$. Fix a complex vector space
$\fh$ of dimension $2n-l$ and its dual $\fh^{*}$, select vectors $ h_{1},
h_{2}, \dots, h_{n} \in \fh$ and $\alpha_{1},\dots , \alpha_{n} \in \fh^{*} $ so
that $ \alpha_{i}(h_{i}) = a_{ij}$.

Let $I= \{i_{1},\dots, i_{n}\} \subset (\Zee/2\Zee)^{n}$; consider the
free Lie superalgebra $ \widetilde{ \fg } (A, I)$ with generators
$e_{1},\dots ,e_{n}$, $f_{1}, \dots, f_{n}$ and $h_{1},\, h_{2},\dots , h_{n}$, where $p(h_{i})=
\ev ,\; p(e_{j})=p(f_{j})=i_{j}$ and defining relations: 
$$
[e_{i}, f_{j}] = \delta_{ij} h_{j};\; [h, e_{i}]=\alpha_{i} (h) e_{i};\;
[h,f_{i}]= -\alpha_{i}(h)f_{i};\; \;[\fh, \fh]=0. \eqno{(1.1)}
$$
Let
$$
Q= \sum_{1 \leq i \leq n} \Zee \alpha_{i};\; \; Q^{\pm} =\{ \alpha \in
\fh  ^{*};\; \alpha = \sum \pm n_{i} \alpha_{i}, n_{i} \in \Zee_{+}\}.
$$

For $\alpha = \sum n_{i} \alpha_{i} \in Q $, set $ht(\alpha) =\sum n_{i}$.
We call $Q$ the set {\it weights} and its subsets $Q^{\pm}$ the sets of
{\it positive} or {\it negative} weights, respectively; $ht(\alpha)$ is
the {\it height} of the weight $\alpha$.

\begin{Statement} {\em ([K1], [vdL]) a)} Let $\tilde\fn_{+}$ and
$\tilde\fn_{-}$ be the superalgebras in $\tilde\fg(A, 
I)$ generated by $ e_{1}$,\dots , $e_{n}$ and $ f_{1},\dots,
f_{n}$, respectively; then $\tilde\fn_{+}$ and
$\tilde\fn_{-}$ are free superalgebras with generators $ e_{1},
\dots , e_{n}$ and $f_{1},\dots, f_{n}$, respectively, and
$\tilde\fg(A, I) \cong \tilde\fn_{+} \oplus \fh\oplus
\tilde\fn_{-}$, as vector superspaces.

{\em b)} Among the ideals of $\tilde\fg (A, I)$ with zero
intersection with $\fh$ there exists a maximal ideal $\fr$
such that $\fr =\fr \bigcap \tilde\fn_{+} \oplus \fr
\bigcap \tilde\fn_{-}$ is the direct sum of ideals.
\end{Statement}

Set $\fg(A, I) = \tilde\fg (A, I)/ \fr$. Neither $\fg(A, I)$ nor $\fg(A, I)' =
[\fg (A, I), \fg(A, I)]$ are simple. As proved in [vdL], the centers $\fc$ of
$ \fg(A, I)$ and $\fc'$ of $ \fg (A, I)'$ consist of all $ h \in
\fh$ such that $ \alpha_{i}(h) =0$ for all
$i= 1,\dots , n$ and the quotient of $\fg(A, I)'$ modulo the center is simple. 

For the symmetrizable matrices $A$ the simple Lie superalgebras of polynomial growth are listed in
[vdL] (they are twisted loop superalgebras); for nonsymmetrizable ones Serganova proved (1989,
unpublished) that these are only $\fpsq(n)^{(2)}$ and a stringy superalgebra
$\fsvect^L_\alpha (1|2)$. (Notice that there is a crucial difference between loop algebras and
stringy algebras: in the former every root vector acts locally nilpotently ([K1]); this is false for
the latter.)

Clearly, the rescaling ($e_i\mapsto\sqrt{\lambda_i}e_i$,
$f_i\mapsto\sqrt{\lambda_i}f_i$) sends $A$ to
$\diag(\lambda_1, \dots , \lambda_n)\cdot A$. Two pairs
$(A, I)$ and $(A', I')$ are said to be {\it equivalent} if
$(A', I')$ is obtained from $(A, I)$ by a permutation of indices or if $A' =
\diag (\lambda_{1}, \dots, \lambda_{n})\cdot A$. Clearly, equivalent pairs
determine isomorphic Lie superalgebra. 

The matrix $A$ (more precisely, a pair $(A, I)$) is called a {\it Cartan matrix} of the Lie
superalgebra $\fg (A, I) $ and also of $\tilde\fg(A, I)$, $\fg'(A, I)$ as well as of $\fg(A,
I)/\fc$ and $\fg' (A, I)/ \fc'$.

Let $\fg$ be one of the Lie superalgebras $\fg'(A, I)$, $\fg(A, I)/\fc$ or $\fg' (A, I)/ \fc'$. Set: 
$$ 
\fg_{\alpha} = \{ g \in \fg :\, [h, g] =
\alpha (h)g \; \text{for any}\; h \in \fh \}
$$
and define the subalgebras $\fn_{\pm}$ of $\fg$ similarly to
$\tilde{\fn}_{\pm}$.

A vector $\alpha \in Q$ is called a {\it root} of $\fg$ if $\fg_{\alpha}
\neq \{0\}$. Denote by $R$ the set of all the roots of $\fg$ and let $R^{\pm} =R
\bigcap Q^{\pm}$. In $R$, introduce a {\it parity}\index{parity of roots} setting:
$$
p(\alpha) = \sum n_{j}i_{j},\; \; \text{where}\; \; \alpha =\sum
n_{j}\alpha_{j} \in R.
$$

\begin{Statement} {\em ([vdL])} Every Lie superalgebra $\fg(A)$ possesses a
root decomposition $\fg = \oplus _{\alpha \in R}\;\fg_{\alpha}$, where
$\fg_{0} =\fh$ and $[\fg_{\alpha}, \fg_{\beta}]\subset \fg_{\alpha +
\beta}$.
\end{Statement}

Therefore, there exists a $\Zee$-grading $\fg = \oplus \fg_{i}$, 
where $\fg_{i} = \bigoplus\limits_{\alpha \in Q\; \text{and}\;
\Ht(\alpha)=i}\fg_{\alpha}$ with $\fn_{\pm} = \oplus _{\alpha \in R^{\pm}}\;
\; \fg_{\alpha}$.

\begin{th*}{Corollary} $\fh$ is a maximal torus of $\fg(A, I) $ and
$\tilde\fg(A, I) $ while $\fh/\fc$ and $\fh/\fc'$ are maximal
tori of $\fg(A, I)/\fc$ and $\tilde\fg (A, I)/\fc'$, respectively.
\end{th*}

Our problem is to describe simple Lie superalgebras of the form $\fg(A, I)$ more
explicitely, i.e., to determine the generators of $\fr$. For an arbitrary $A$ this is an
open problem even for Lie algebras. 

\ssec{1.1. Bases (systems of simple roots)} Let $R$ be the root system of $\fg$.
For any subset $B=\{\sigma_{1},
\dots, \sigma_{n}\} \subset R$ , set:
$$
R_{B}^{\pm} =\{ \alpha \in R \;:\;\; \alpha =
\pm \sum_{\pm} n_{i}
\sigma_{i},\;\;n_{i} \in \Zee_{+} \}.
$$
Clearly, $\dim \fg _{\pm \sigma_{i}} = (1, 0)$ or $(0, 1)$ and $R_{B}^{+}
\bigcap R_{B}^{-} =\{0\}$.

The set $B$ is called a {\it base}\index{base, see simple roots, system of}
\index{simple roots, system of} of $R$ (or
$\fg$) or a {\it system of simple roots} if $ \sigma_{1}, \cdots , \sigma_{n} $ are
linearly independent and there exist $\tilde{e_{1}} \in~\fg _{\sigma_{1}},\dots ,
\tilde{e_{n}}~\in \fg _{\sigma_{n}}$, $\tilde{f_{1}} \in~\fg _ {- \sigma
_{1}},\dots , \tilde{f_{n}} \in \fg _{ -\sigma _{n}}$ such that:
$$
\fg = \oplus \fg _{B}^{-} \oplus \fh 
\oplus \fg _{B}^{+},
$$
where $\fg _{B}^{-}$ (resp. $\fg _{B}^{+}$) is the super\-algebra
gen\-er\-ated by $\tilde{e_{1}},\dots , \tilde{e_{n}}$
 (resp. $\tilde{f_{1}},\dots, \tilde{f_{n}})$.

Let $B$ be a base and $\tilde{e_{1}}, \dots , \tilde{e_{n}}$, $\tilde{f_{1}},
\dots, \tilde{f_{n}}$, the corresponding elements of $\fg$. Set
$\tilde{h_{i}}=[\tilde{e_{i}}, \tilde{f_{i}}]$ , $A_{B} =(a_{ij})$, where
$a_{ij} =\sigma_{i}(\tilde{h_{j}})$ and $I_{B}=\{p(\sigma_{1}), \cdots,
p(\sigma_{n})\}$.

The matrix $A_{B}$ or, more precisely, the pair $(A_{B}, I_{B})$, is called the
{\it Cartan matrix}\index{Cartan matrix}\index{matrix Cartan} of $\fg$. The
elements 
$\tilde{e_{i}}$, $\tilde{f_{i}}$ and $ \tilde{h_{i}}$ for $i=1,\dots,
n$, satisfy the relations (1.1). Since there is no ideal in $\fg$ with
a non-zero intersection with $\fh$, we have: $\fg =\fg(A, I)$.

Two bases $B_{1}$ and $B_{2}$ are called {\it equivalent} if the pairs
$(A_{B_{1}}, I_{B_{1}})$ and $(A_{B_{2}}, I_{B_{2}})$ are equivalent.

Hereafter $\fg =\fg (A, I)$. How many Cartan matrices correspond to the same Lie
superalgebra $\fg$? Let $\fg(A)$ be a Lie superalgebra with Cartan matrix.

The following proposition due to V. Serganova lists, up to equivalence, all bases of
$\fg$ and, therefore, all Cartan matrices.

\begin{Proposition} Let $B$ be a base, $\tilde{e_{i}}$, $\tilde{f_{i}}$,
for $i=1,\dots, n$ the corresponding set of
generators and $A_{B}=(a_{ij})$ the Cartan matrix. Fix an $i$. Then:

{\em a)} If $p(\sigma_{i})= \ev $ then, if $\fg$ is of polynomial growth, $a_{ii}
\neq 0$ and the Lie subalgebra generated by the $e_{i}$ and $f_{i}$ is
isomorphic to $\fsl(2)$.

{\em b)} If $p(\sigma_{i}) =\od $ and $a_{ii}=0$, then $\sigma_{i} \not \in R$
and the subsuperalgebra generated by the $e_{i}$ and $f_{i}$ is isomorphic to
$\fsl(1|1)$.

{\em c)} If $p(\sigma_{i}) =\od $ and $a_{ii} \neq 0$, then $\sigma_{i} \not \in R$
and
the subsuperalgebra generated by the $e_{i}$ and $f_{i}$ is isomorphic to
$\fosp(1|2)$.
\end{Proposition}

\ssec{1.3. Chevalley generators and odd reflections}
Let us multiply $A_{B}$ from the right by a diagonal matrix so that in the
cases a), b) or c) of Proposition 1.2 the diagonal elements of $A_{B}$ become 2, 0
or 1, respectively. Such a matrix is said to be {\it normed}.

\begin{Convention} In what
follows we only consider normed matrices.
\end{Convention} 

A typical way to represent Lie algebras with integer Cartan matrices is via graphs
called in the finite dimensional case Dynkin diagrams. The Cartan matrices of Lie
superalgebras can be nonsymmetrizable or have complex entries; hence it is not always
possible to assign to them an analog of the Dynkin diagram. 

Every integer Cartan matrix $(A, I)$ can be encoded with an analog of
Dynkin diagram. Namely, the {\it Dynkin--Kac diagram}\index{Dynkin--Kac diagram} of
the matrix $(A, I)$ is the set of $n$ nodes (vertices) connected by multiple edges, perhaps
endowed with an arrow, according to the following rules. The nodes are of four
types:

{\sl To every simple root there corresponds}
$$
\left\{\begin{array}{l}
\text{a vertex }\; \circ\; \text{ if }\; p(\alpha_{i})= \ev ,\; \text{ and }\;
a_{ii}=2\\
\text{a vertex}\; \otimes\text{ if }\; p(\alpha_{i}) =\od \; \text{ and }\; 
a_{ii}=0,\\
\text{a vertex }\; \bullet \text{ if }\;
p(\alpha_{i}) =\od \; \text{ and }\; a_{ii}=1;\\
\text{a vertex }\; *\; \text{ if }\; p(\alpha_{i})= \ev ,\; \text{ and }\;
a_{ii}=0.\\
\end{array}\right.
$$

{\it A posteriori} we find out that the roots $*$ can only occure if 
$\fg(A, I)$ grows faster than polynomially. 

Let the nodes corresponding to the $i$-th and the $j$-th roots be connected with 
$\max(|a_{ij}|, |a_{ji}|)$ edges andowed with the sign $>$ pointing to the
$j$-th node if $|a_{ij}| >|a_{ji}|$. 

$\bullet$ It turns out that an integer Cartan matrix $(a_{ij})$ and a sequence $I
=\{ i_{1},\dots, i_{n}\}$ connected with a base can be uniquely, up to
equivalence, recovered from their Dynkin
diagram in all cases except $\fg =\fd(\alpha)$, $\fg =
\fd(\alpha)^{(1)}$ or $\fsl(2|4)^{(2)}$. The procedure is as follows:

1) If the $i$-th and the $j$-th nodes are connected by $k$ segments with an
arrow pointing towards the $j$-th node, set:
$$
|a_{ij}|=k, |a_{ji}|=1;
$$

2) If the $i$-th and the $j$-th nodes
are joined by $k$ segments without
arrows, set
$$ 
|a_{ij}|=|a_{ji}|=k.
$$

$$ \begin{array}{lll}
\text{If the $j$-th node is $\bullet$, then } & a_{jj}=1 & i_{j}=\od, \\ 
\text{If the $j$-th node is $\otimes$, then} & a_{jj}=0 & i_{j} =\od, \\ 
\text{If the $j$-th node is $\circ$, then} & a_{jj}=2 & i_{j}=\ev. 
\end{array}
\leqno{\text{3)}}
$$

$$ 
\text{If $a_{ii} \neq 0$ then}\; \; a_{ji} \leq 0\; \text{for any $j \neq i$.}
\leqno{\text{4)}} 
$$

5) If $a_{ij}=0$, then the $i$-th column is recovered, up to multiplication by
$-1$, as follows: if $a_{ji} \neq 0$ for exactly one $j$,
then the sign of $a_{ji}$ may be
chosen arbitrarily.

If $a_{j_{1}i}$, $a_{j_{2}i} \neq 0$ for exactly two distinct indices
$j_{1}$, $j_{2}$, then $ a_{j_{1}i}/ a_{j_{2}i} <0$.

If $a_{j_{1}i}$, $a_{j_{2}i}$, $a_{j_{3}i} \neq 0$ for exactly three
distinct indices $j_{1}$, $j_{2}$, $j_{3}$ so that $a_{j_{1}j_{2}} \neq 0$,
$a_{j_{1}j_{3}}=a_{j_{2}j_{3}}= 0$, then $a_{j_{1}i}/a_{j_{2}i}>0$, 
$a_{j_{1}i}/a_{j_{3}i} <0$.

The set of generators corresponding to a
normed matrix is often denoted in what follows by $X_{1}^{+},\dots , X_{n}^{+}$ and
$X_{1}^{-}, \dots , X_{n}^{-}$ instead of $e_{1},\dots , e_{n}$, and $f_{1}, \dots, f_{n}$,
respectively, and $X_{i}^{\pm}$, $1 \leq i \leq n$ are called the {\it
Chevalley generators}. 

The reflection in the $i$th root sends one set of Chevalley generators into the new one:
$\tilde X_{i}^{\pm}=X_{i}^{\mp}$; $\tilde X_{j}^{\pm}=[X_{i}^{\pm}, X_{j}^{\pm}]$ if
$a_{ij}\neq 0$ and $\tilde X_{j}^{\pm}=X_{j}^{\pm}$ otherwise. The reflections in roots with
$a_{ii}\neq 0$ generate the Weyl group of $\fg_\ev$. For the discussion of what is generated by the
other reflections, called {\it odd} ones, see [LSS], [S] and [E]. It is instructive to compare [E]
with [PS].

\ssec{1.4. Serre-type relations} Let $\fg=\fg(A)$.
Let $\fh \subset \fg$ be a maximal torus (i.e., the maximal diagonalizing
commutative subalgebra), $\fg^{+}$, $\fg^{-}$ subalgebras of
$\fg$ generated by root vectors corresponding to positive (resp. negative) roots; 
let the rank $\rk \fg$ of $\fg$ be equal to $n=\dim \fg$; let $X_{i}^{\pm}$, where
$1\leq i \leq n$, be root
vectors corresponding to simple roots (for $+$) and their opposite (for $-$). Set
$H_{i}=[X_{i}^{+}, X_{i}^{-}].$ It is subject to a direct verification that
$$
[H_{i}, H_{i}]=0, \;\;[X_{i}^{+}, X_{j}^{-}]=
\delta_{ij}H_{i}, \;\;
[H_{i}, X_{j}^{\pm}]=\pm a_{ij}X_{j}^{\pm}
\eqno{(SR_{0})}
$$

Clearly, the generators of
$\fn^{\pm}$ are $X_{1}^{\pm}, X_{2}^{\pm}, \dots , 
X_{n}^{\pm}$. The defining relations are found by induction on
$n$ with the help of
the Hochschild--Serre spectral sequence (see [Fu], [GM]). For the basis
of the induction consider the following cases:
$$
\circ \;\; or\;\; \bullet :\;\;\text{no relations}
;\;\;\; {\otimes}:
[X^{\pm}, X^{\pm}]=0. \eqno{(1.4.1)}
$$
Set $\deg X_{i}^{\pm} = 0$
for $ 1\leq i\leq n-1$ and $\deg X_{n}^{\pm} =
\pm 1$. Let $ \fn^{\pm} = \oplus \fn_{ i}^{\pm}$, 
$\fg= \oplus \fg_{i} $ be the
corresponding $\Zee $-gradings. From the Hochschield--Serre spectral sequence for
the pair $\fn_{ 0}^{\pm} \subset \fn ^{\pm}$ we get (with $\fn_{\pm}
=\fn^{\pm}/\fn_{ 0}^{\pm}$):
$$
H_{2}(\fn_{0}^{\pm})\oplus H_{1}(\fn_{
0}^{\pm}; H_{ 1}(\fn_{\pm}))\oplus H_{0}(\fn_
{ 0}^{\pm}; H_{
2}(\fn_{\pm})). 
\eqno{(1.4.2)}
$$
In the cases we are considering it is
clear that
$$
H_{1}(\fn_{\pm})= \fn_{ 1}^{\pm} , \;\;\;
H_{ 2}(\fn_{\pm}) = E^{2}(\fn_
{ 1}^{\pm})/\fn_{ 2}^{\pm}
\eqno{(1.4.3)}
$$
and, therefore, the second summand in (1.4.2)
provides us with relations of the form:
$$
(ad~X_{n}^{\pm})^{k_{ni}} (X_{i}^{\pm})=0
\;\;\;\text{if the $n$-th root is not}\;\;
{\otimes}
$$
or
$$
[X_{n}, X_{n}]=0 \;\;\;\text{if the $n$-th root is}
\;\;\otimes.
$$
while the third summand in (1.4.2) consists of $\fn_{0}^{\pm}$-lowest 
vectors in
$$
E^{2}(\fn_{1}^{\pm})/(\fn_{ 2}^{\pm} +
\fn^{\pm} 
E^{2}(\fn_{1}^{\pm})).
$$

Let the matrix $B=(b_{ij})$ be obtained from the Cartan matrix $A=(a_{ij})$
by replacing all nonzero elements in the row with $a_{ii}=0$ by $-1$
and multiplying the row
with $a_{ii}=1$ by $2$. The following proposition is
straightforward: 

\begin{Proposition} The numbers $k_{in}$ and $k_{ni}$
are expressed in terms
of $(b_{ij})$ as follows:
$$
\matrix (\ad~X_{i}^{\pm})^{1-b_{ij}}(X_{j}
^{\pm})=0 &
\text{ for $ i \neq j$} \\ & \\
\text{$[X_{i}^{\pm}, X_{i}^{\pm}]=0$} & \text{if
$a_{ii} =0$} \endmatrix \eqno{(SR_{\pm})}
$$
\end{Proposition}
The relations $(SR_{0})$ and $(SR_{\pm})$ will be called Serre relations for Lie
superalgebra $\fg(A)$.

\ssec{1.5. Non-Serre-type relations} Let us consider the simplest case: $\fsl
(m|n)$ in the realization with the base $\bigcirc --\cdots --\bigcirc
--\otimes --\bigcirc --\cdots \bigcirc$. Then
$H_2(\fn_{\pm})$ from the third summand in (2.2) is just $E^2(\fn_{\pm})$. 

Let us confine ourselves to the positive roots for simplicity. Let $X_{1}$, 
\dots , $X_{m-1}$; $Y_{1}$, \dots , $Y_{n-1}$ be the root vectors corresponding
to even roots, $Z$ the root vector corresponding to the root $\otimes$.

If $n=1$ or $m=1$, then $E^2(\fn)$ is an irreducible $\fn_{\bar 0}$-module
and there are no non-Serre relations. If $n\neq 1$ and $m\neq 1$, then
$E^2(\fn)$ splits into 2 irreducible $\fn_{\bar 0}$-modules. The lowest
component of one of them corresponds to the relation $[Z, Z]=0$, the other
one corresponds to the non-Serre-type relation 
$$ 
[[X_{m-1}, Z], [Y_{1}, Z]]
=0. \eqno{(*)} 
$$

If instead of $\fsl (m|n)$ we would have considered the Lie algebra $\fsl(m+n)$
the same argument would have led us to the two relations: $[Z, [Z, 
X_{m-1}]]=0$ and $[Z, [Z, Y_{1}]]=0$ both of Serre type.

\vskip 0.2 cm

Let us consider the other root systems for the simplest example
$\fsl(1|n)$ to see what might happen. We start from the simplest base (one
grey root) and apply to it odd reflections, see [PS], with respect to the first and then
second root. We get the generators as indicated that satisfy, besides Serre relations, the
relations indicated: 
$$
\begin{tabular}{ccl} 
diagram&the corresponding generators& non-Serre relations\cr
$\otimes-\circ-\circ-\circ-\circ-$&$X_1$, $X_2$, $X_3$, $X_4$, $X_5$&\cr 
$\otimes-\otimes-\circ-\circ-\circ-$&$X^-_1$, $[X_1, X_2]$, $X_3$, $X_4$, $X_5$&$[[X_1, 
X_2], X_4]=0$\cr 
$\circ-\otimes-\otimes-\circ-\circ-$&$X_2$, $[X^-_1, X^-_2]$, $[[X_1, X_2], X_3]$, 
$X_4$, $X_5$&$[[[X_1, X_2], X_3], X_2]=0$, $[[X^-_1, X^-_2], X_4]=0$, \cr
&&$[[[X_1, X_2], X_3], X_5]=0$\cr 
\end{tabular}
$$
For $\fsl(m+n)$ we similarly have (the passage from diagram to diagram is given
by odd reflections in the 3rd, 4th, 2nd roots, respectively:
$$
\begin{tabular}{ccl} 
diagram&the corresponding generators& non-Serre relations\cr
$\circ-\circ-\otimes-\circ-\circ-$&$X_1$, $X_2$, $X_3$, $X_4$, $X_5$&\cr 
$\circ-\otimes-\otimes-\otimes-\circ-$&$X_1$, $[X_3, X_2]$, $X^-_3$, $[X_3, 
X_4]$, $X_5$&$[[X_3, X_2], [X_3, X_4]]=0$\cr 
$\circ-\otimes-\circ-\otimes-\otimes-$&$X_1$, $[X_3, X_2]$, $X_4$, $[X^-_3, 
X^-_4]$,&$[[[X_1, X_2], X_3], X_2]=0, \dots$\cr 
& $[[X_3, X_4], X_5]$&\cr
$\otimes-\otimes-\otimes-\otimes-\otimes-$&$[X_1, [X_3, X_2]]$, $[X^-_3, X^-_2]$, 
&$[[X_1, [X_2, X_3]], [[X_3,
X_4], X_5]]=0, \dots$\cr 
&$[[X_3, X_2], X_4]$, $[X^-_3, X^-_4]$, $[[X_3, X_4], X_5]$&\cr
\end{tabular}
$$
The idea of construction of the relations is clear, so we do not list all of them but
leave the completion of the third column as an exersise. Contrarywise it is absolutely
unclear, how to single out the {\it basic} relations. In what follows we list all the
basic relations for the exceptional Lie superalgebras and for the \lq\lq key cases" for the
remaining series. The relations listed in [LSS] and [LSe], namely, all the Serre
relations for {\it all} bases are, clearly, very redundant.

\begin{Problem} How to single out the basic relations in the general case?
How to describe the change of relations under the action of odd reflections?
\end{Problem}

\ssbegin{1.6}{Theorem} All the non-Serre-type relations become Serre relations
after an appropriate odd reflection from a super Weyl group. {\em (For the 
definition of a super Weyl group see [LSS], [S] and [E].) }
\end{Theorem}

In Tables below for a finite dimensional algebra its dimension is indicated. The
generators are assumed to correspond to positive roots. Their parities are determined
by the corresponding diagonal elements of the Cartan matrix, since we do not consider the algebras
of infinite growth. We only consider the Chevalley generators corresponding to the positive
roots.

\section*{\S 2. Table 2.1. Relations for symmetrizable Cartan matrices}

\noindent
$\fsl(2|2)$, dim =
$(7|8)$

$$
\begin{pmatrix}
2 & -1 & 0 \\ -1 & 0 & 1 \\ 0 & -1 & 2
\end{pmatrix}\quad \begin{matrix}
{{[x_{1}, x_{3}]} = 0}\\
{{[x_{2}, x_{2}]} = 0}\\
{{[x_{1}, [x_{1}, x_{2}]]} = 0}\\
{{[x_{3}, [x_{2}, x_{3}]]} = 0} \\
{{[[x_{1}, x_{2}], [x_{2}, x_{3}]]} = 0}
\end{matrix}\quad\quad \begin{pmatrix}
0 & 1 & 0 \\ -1 & 0 & 1 \\ 0 & -1 & 0
\end{pmatrix}\quad \begin{matrix}
{{[x_{1}, x_{1}]} = 0} \\
{{[x_{1}, x_{3}]} = 0} \\
{{[x_{2}, x_{2}]} = 0} \\
{{[x_{3}, x_{3}]} = 0} \\
{{[[x_{1}, x_{2}], [x_{2}, x_{3}]]} = 0}
\end{matrix}
$$

$----------------------------------------$

\noindent $\fsl(1|3)$, dim = $(9|6)$

$$
\begin{pmatrix}
0 & 1 & 0 \\ -1 & 2 & -1 \\ 0 & -1 & 2
\end{pmatrix}\quad \begin{matrix}
{{[x_{1}, x_{1}]} = 0} \\
{{[x_{1}, x_{3}]} = 0} \\
{{[x_{2}, [x_{1}, x_{2}]]} = 0} \\
{{[x_{2}, [x_{2}, x_{3}]]} = 0} \\
{{[x_{3}, [x_{2}, x_{3}]]} = 0}
\end{matrix}\quad\quad \begin{pmatrix}
0 & 1 & 0 \\ -1 & 0 & 1 \\ 0 & -1 & 2
\end{pmatrix}\quad \begin{matrix}
{{[x_{1}, x_{1}]} = 0} \\
{{[x_{1}, x_{3}]} = 0} \\
{{[x_{2}, x_{2}]} = 0} \\
{{[x_{3}, [x_{2}, x_{3}]]} = 0} \\
{{[[x_{1}, x_{2}], [x_{2}, x_{3}]]} = 0}
\end{matrix}
$$

$----------------------------------------$
\noindent $\fa\fg_2$, dim = $(17|14)$

$$
\begin{pmatrix}
0 & 1 & 0 \\ -1 & 2 & -3 \\ 0 & -1 & 2
\end{pmatrix}\quad \begin{matrix}
{{[x_{1}, x_{1}]} = 0} \\
{{[x_{1}, x_{3}]} = 0} \\
{{[x_{2}, [x_{1}, x_{2}]]} = 0} \\
{{[x_{3}, [x_{2}, x_{3}]]} = 0} \\
{{[x_{2}, [x_{2}, [x_{2}, [x_{2}, x_{3}]]]]} = 0}
\end{matrix}\quad\quad \begin{pmatrix}
0 & 1 & 0 \\ -1 & 0 & 3 \\ 0 & -1 & 2
\end{pmatrix}\quad \begin{matrix}
{{[x_{1}, x_{1}]} = 0} \\
{{[x_{1}, x_{3}]} = 0} \\
{{[x_{2}, x_{2}]} = 0} \\
{{[x_{3}, [x_{2}, x_{3}]]} = 0} \\
{{[[x_{1}, x_{2}], [[x_{1}, x_{2}], [[x_{1}, x_{2}], [x_{2}, x_{3}]]]]} = 0}
\end{matrix}
$$
$$
\begin{pmatrix}
0 & -3 & 1 \\ -3 & 0 & 2 \\ -1 & -2 & 2
\end{pmatrix}\quad \begin{matrix}
{{[x_{1}, x_{1}]} = 0} \\
{{[x_{2}, x_{2}]} = 0} \\
{{[x_{2}, [x_{1}, x_{3}]]} = {{-\frac{1}{3}[x_{3}, [x_{1}, x_{2}]]}}} \\
{{[x_{3}, [x_{1}, x_{3}]]} = 0} \\
{{[x_{3}, [x_{3}, [x_{2}, x_{3}]]]} = 0}
\end{matrix}
$$
$$
\begin{pmatrix}
2 & -1 & 0 \\ -3 & 0 & 2 \\ 0 & -1 & 1
\end{pmatrix}\quad \begin{matrix}
{{[x_{1}, x_{3}]} = 0} \\
{{[x_{2}, x_{2}]} = 0} \\
{{[x_{1}, [x_{1}, x_{2}]]} = 0} \\
{{[[x_{2}, x_{3}], [x_{3}, x_{3}]]} = 0} \\
{{[x_{3}, [[x_{1}, x_{2}], [x_{2}, x_{3}]]]} = 
{{-\frac{1}{2}[[x_{2}, x_{3}], [x_{3}, [x_{1}, x_{2}]]]}}}
\end{matrix}
$$

$----------------------------------------$

\noindent $\fosp(3|2)$, dim = $(6|6)$
$$
\begin{pmatrix}
0 & 1 \\ -2 & 2
\end{pmatrix}\quad \begin{matrix}
{{[x_{1}, x_{1}]} = 0} \\
{{[x_{2}, [x_{2}, [x_{1}, x_{2}]]]} = 0}
\end{matrix}\quad\quad \begin{pmatrix}
0 & 1 \\ -1 & 1
\end{pmatrix}\quad \begin{matrix}
{{[x_{1}, x_{1}]} = 0} \\
{{[[x_{1}, x_{2}], [x_{2}, x_{2}]]} = 0}
\end{matrix}
$$

$----------------------------------------$

\noindent $\fosp(2|4)$, dim = $(11|8)$
$$
\begin{pmatrix}
0 & 1 & 0 \\ -1 & 2 & -2 \\ 0 & -1 & 2
\end{pmatrix}\quad \begin{matrix}
{{[x_{1}, x_{1}]} = 0} \\
{{[x_{1}, x_{3}]} = 0} \\
{{[x_{2}, [x_{1}, x_{2}]]} = 0} \\
{{[x_{3}, [x_{2}, x_{3}]]} = 0} \\
{{[x_{2}, [x_{2}, [x_{2}, x_{3}]]]} = 0}
\end{matrix}
\quad\quad \begin{pmatrix}
0 & 1 & 0 \\ -1 & 0 & 2 \\ 0 & -1 & 2
\end{pmatrix}\quad \begin{matrix}
{{[x_{1}, x_{1}]} = 0} \\
{{[x_{1}, x_{3}]} = 0} \\
{{[x_{2}, x_{2}]} = 0} \\
{{[x_{3}, [x_{2}, x_{3}]]} = 0} \\
{{[[x_{1}, x_{2}], [[x_{1}, x_{2}], [x_{2}, x_{3}]]]} = 0}
\end{matrix}
$$
$$\begin{pmatrix}
0 & -2 & 1 \\ -2 & 0 & 1 \\ -1 & -1 & 2
\end{pmatrix}\quad \begin{matrix}
{{[x_{1}, x_{1}]} = 0} \\
{{[x_{2}, x_{2}]} = 0} \\
{{[x_{2}, [x_{1}, x_{3}]]} = {{-\frac{1}{2}[x_{3}, [x_{1}, x_{2}]]}}} \\
{{[x_{3}, [x_{1}, x_{3}]]} = 0} \\
{{[x_{3}, [x_{2}, x_{3}]]} = 0}
\end{matrix}
$$

$----------------------------------------$

\noindent $\fosp(3|4)$, dim = $(5|6)$
$$
\begin{pmatrix}
2 & -1 & 0 \\ -1 & 0 & 1 \\ 0 & -2 & 2
\end{pmatrix}\quad \begin{matrix}
{{[x_{1}, x_{3}]} = 0} \\
{{[x_{2}, x_{2}]} = 0} \\
{{[x_{1}, [x_{1}, x_{2}]]} = 0} \\
{{[x_{3}, [x_{3}, [x_{2}, x_{3}]]]} = 0} \\
{{[[x_{1}, x_{2}], [x_{2}, x_{3}]]} = 0}
\end{matrix}\quad\quad \begin{pmatrix}
0 & 1 & 0 \\ -1 & 0 & 1 \\ 0 & -1 & 1
\end{pmatrix}\quad \begin{matrix}
{{[x_{1}, x_{1}]} = 0} \\
{{[x_{1}, x_{3}]} = 0} \\
{{[x_{2}, x_{2}]} = 0} \\
{{[[x_{1}, x_{2}], [x_{2}, x_{3}]]} = 0} \\
{{[[x_{2}, x_{3}], [x_{3}, x_{3}]]} = 0}
\end{matrix}
$$
$$
\begin{pmatrix}
0 & 1 & 0 \\ -1 & 2 & -1 \\ 0 & -1 & 1
\end{pmatrix}\quad \begin{matrix}
{{[x_{1}, x_{1}]} = 0} \\
{{[x_{1}, x_{3}]} = 0} \\
{{[x_{2}, [x_{1}, x_{2}]]} = 0} \\
{{[x_{2}, [x_{2}, x_{3}]]} = 0} \\
{{[[x_{2}, x_{3}], [x_{3}, x_{3}]]} = 0}
\end{matrix}
$$

$----------------------------------------$

\noindent $\fosp(5|2)$, dim $= (13|10)$

$$
\begin{pmatrix}
2 & -1 & 0 \\ -1 & 0 & 1 \\ 0 & -1 & 1
\end{pmatrix}\quad \begin{matrix}
{{[x_{1}, x_{3}]} = 0} \\
{{[x_{2}, x_{2}]} = 0} \\
{{[x_{1}, [x_{1}, x_{2}]]} = 0} \\
{{[[x_{1}, x_{2}], [x_{2}, x_{3}]]} = 0} \\
{{[[x_{2}, x_{3}], [x_{3}, x_{3}]]} = 0}
\end{matrix}\quad\quad \begin{pmatrix}
0 & 1 & 0 \\ -1 & 0 & 1 \\ 0 & -2 & 2
\end{pmatrix}\quad \begin{matrix}
{{[x_{1}, x_{1}]} = 0} \\
{{[x_{1}, x_{3}]} = 0} \\
{{[x_{2}, x_{2}]} = 0} \\
{{[x_{3}, [x_{3}, [x_{2}, x_{3}]]]} = 0} \\
{{[[x_{1}, x_{2}], [x_{2}, x_{3}]]} = 0}
\end{matrix}
$$
$$
\begin{pmatrix}
0 & 1 & 0 \\ -1 & 2 & -1 \\ 0 & -2 & 2
\end{pmatrix}\quad \begin{matrix}
{{[x_{1}, x_{1}]} = 0} \\
{{[x_{1}, x_{3}]} = 0} \\
{{[x_{2}, [x_{1}, x_{2}]]} = 0} \\
{{[x_{2}, [x_{2}, x_{3}]]} = 0} \\
{{[x_{3}, [x_{3}, [x_{2}, x_{3}]]]} = 0}
\end{matrix}
$$

$----------------------------------------$

\noindent $\fosp_\alpha(4|2)$, dim $= (9|8)$

$$
\begin{pmatrix}
2 & -1 & 0 \\ \alpha & 0 & \ -1 - \alpha \\ 0 & -1 & 2
\end{pmatrix}\quad \begin{matrix}
{{[x_{1}, x_{3}]} = 0} \\
{{[x_{2}, x_{2}]} = 0} \\
{{[x_{1}, [x_{1}, x_{2}]]} = 0} \\
{{[x_{3}, [x_{2}, x_{3}]]} = 0}
\end{matrix}\quad\quad \begin{pmatrix}
2 & -1 & 0 \\ -1 & 0 & -\alpha \\ 0 & -1 & 2
\end{pmatrix}\quad \begin{matrix}
{{[x_{1}, x_{3}]} = 0} \\
{{[x_{2}, x_{2}]} = 0} \\
{{[x_{1}, [x_{1}, x_{2}]]} = 0} \\
{{[x_{3}, [x_{2}, x_{3}]]} = 0}
\end{matrix}
$$
$$
\begin{pmatrix}
0 & 1 & -1 - \alpha \\ 
-1 & 0 & -\alpha \\ 
 -1 - \alpha & \alpha & 0
\end{pmatrix}\quad \begin{matrix}
{{[x_{1}, x_{1}]} = 0} \\
{{[x_{2}, x_{2}]} = 0} \\
{{[x_{3}, x_{3}]} = 0} \\
{{[x_{2}, [x_{1}, x_{3}]]} = 
{\left( -1 - \alpha \right) \, [x_{3}, [x_{1}, x_{2}]]}}
\end{matrix}
$$
$$
\begin{pmatrix}
0 & 1 & -1 - \alpha \\ 
-1 & 0 & -\alpha \\ 
 -1 - \alpha & \alpha & 0
\end{pmatrix}\quad \begin{matrix}
{{[x_{1}, x_{1}]} = 0} \\
{{[x_{2}, x_{2}]} = 0} \\
{{[x_{3}, x_{3}]} = 0} \\
{{[x_{2}, [x_{1}, x_{3}]]} = 
{\left( -1 - \alpha \right) \, [x_{3}, [x_{1}, x_{2}]]}}
\end{matrix}\quad\quad \begin{pmatrix}
2 & -1 & 0 \\ -1 & 0 & 1+\alpha \\ 0 & -1 & 2
\end{pmatrix}\quad \begin{matrix}
{{[x_{1}, x_{3}]} = 0} \\
{{[x_{2}, x_{2}]} = 0} \\
{{[x_{1}, [x_{1}, x_{2}]]} = 0} \\
{{[x_{3}, [x_{2}, x_{3}]]} = 0}
\end{matrix}
$$

$----------------------------------------$

\noindent $\fosp(6|2)$, dim $= (18|12)$

$$
\begin{pmatrix}
0 & 1 & 0 & 0 \\ -1 & 2 & -1 & -1 \\ 0 & -1 & 2 & 0 \\ 
 0 & -1 & 0 & 2
\end{pmatrix}\quad \begin{matrix}
{{[x_{1}, x_{1}]} = 0} \\
{{[x_{1}, x_{3}]} = 0} \\
{{[x_{1}, x_{4}]} = 0} \\
{{[x_{3}, x_{4}]} = 0} \\
{{[x_{2}, [x_{1}, x_{2}]]} = 0} \\
{{[x_{2}, [x_{2}, x_{3}]]} = 0} \\
{{[x_{2}, [x_{2}, x_{4}]]} = 0} \\
{{[x_{3}, [x_{2}, x_{3}]]} = 0} \\
{{[x_{4}, [x_{2}, x_{4}]]} = 0}
\end{matrix}\quad\quad \begin{pmatrix}
0 & 1 & 0 & 0 \\ -1 & 0 & 1 & 1 \\ 0 & -1 & 2 & 0 \\ 0 & -1 & 0 & 2
\end{pmatrix}\quad \begin{matrix}
{{[x_{1}, x_{1}]} = 0} \\
{{[x_{1}, x_{3}]} = 0} \\
{{[x_{1}, x_{4}]} = 0} \\
{{[x_{2}, x_{2}]} = 0} \\
{{[x_{3}, x_{4}]} = 0} \\
{{[x_{3}, [x_{2}, x_{3}]]} = 0} \\
{{[x_{4}, [x_{2}, x_{4}]]} = 0} \\
{{[[x_{1}, x_{2}], [x_{2}, x_{3}]]} = 0} \\
{{[[x_{1}, x_{2}], [x_{2}, x_{4}]]} = 0}
\end{matrix}
$$
$$\begin{pmatrix}
2 & -1 & 0 & 0 \\ -1 & 0 & 1 & 1 \\ 0 & 1 & 0 & -2 \\ 
 0 & 1 & -2 & 0
\end{pmatrix}\quad \begin{matrix}
{{[x_{1}, x_{3}]} = 0} \\
{{[x_{1}, x_{4}]} = 0} \\
{{[x_{2}, x_{2}]} = 0} \\
{{[x_{3}, x_{3}]} = 0} \\
{{[x_{4}, x_{4}]} = 0} \\
{{[x_{1}, [x_{1}, x_{2}]]} = 0} \\
{{[x_{3}, [x_{2}, x_{4}]]} = {[x_{4}, [x_{2}, x_{3}]]}} \\
{{[[x_{1}, x_{2}], [x_{2}, x_{3}]]} = 0} \\
{{[[x_{1}, x_{2}], [x_{2}, x_{4}]]} = 0}
\end{matrix}
$$

$----------------------------------------$

\noindent $\fosp(4|4)$, dim $= (16|16)$

$$
\begin{pmatrix}
2 & -1 & 0 & 0 \\ -1 & 0 & 1 & 1 \\ 0 & -1 & 2 & 0 \\ 
 0 & -1 & 0 & 2
\end{pmatrix}\quad \begin{matrix}
{{[x_{1}, x_{3}]} = 0} \\
{{[x_{1}, x_{4}]} = 0} \\
{{[x_{2}, x_{2}]} = 0} \\
{{[x_{3}, x_{4}]} = 0} \\
{{[x_{1}, [x_{1}, x_{2}]]} = 0} \\
{{[x_{3}, [x_{2}, x_{3}]]} = 0} \\
{{[x_{4}, [x_{2}, x_{4}]]} = 0} \\
{{[[x_{1}, x_{2}], [x_{2}, x_{3}]]} = 0} \\
{{[[x_{1}, x_{2}], [x_{2}, x_{4}]]} = 0}
\end{matrix}\quad\quad \begin{pmatrix}
0 & 1 & 0 & 0 \\ -1 & 0 & 1 & 1 \\ 0 & 1 & 0 & -2 \\ 0 & 1 & -2 & 0
\end{pmatrix}\quad \begin{matrix}
{{[x_{1}, x_{1}]} = 0} \\
{{[x_{1}, x_{3}]} = 0} \\
{{[x_{1}, x_{4}]} = 0} \\
{{[x_{2}, x_{2}]} = 0} \\
{{[x_{3}, x_{3}]} = 0} \\
{{[x_{4}, x_{4}]} = 0} \\
{{[x_{3}, [x_{2}, x_{4}]]} = {[x_{4}, [x_{2}, x_{3}]]}} \\
{{[[x_{1}, x_{2}], [x_{2}, x_{3}]]} = 0} \\
{{[[x_{1}, x_{2}], [x_{2}, x_{4}]]} = 0}
\end{matrix}
$$

$----------------------------------------$

\noindent $\fab_3$, dim $= (24|16)$

$$
\begin{pmatrix}
2 & -1 & 0 & 0 \\ -3 & 0 & 1 & 0 \\ 0 & -1 & 2 & -2 \\ 
 0 & 0 & -1 & 2
\end{pmatrix}\quad \begin{matrix}
{{[x_{1}, x_{3}]} = 0} \\
{{[x_{1}, x_{4}]} = 0} \\
{{[x_{2}, x_{2}]} = 0} \\
{{[x_{2}, x_{4}]} = 0} \\
{{[x_{1}, [x_{1}, x_{2}]]} = 0} \\
{{[x_{3}, [x_{2}, x_{3}]]} = 0} \\
{{[x_{4}, [x_{3}, x_{4}]]} = 0} \\
{{[x_{3}, [x_{3}, [x_{3}, x_{4}]]]} = 0} \\
{{[[x_{3}, x_{4}], [[x_{1}, x_{2}], [x_{2}, x_{3}]]]} = 
{2\, [[x_{3}, [x_{1}, x_{2}]], [x_{4}, [x_{2}, x_{3}]]]}}
\end{matrix}
$$
$$
\begin{pmatrix}
0 & -3 & 1 & 0 \\ -3 & 0 & 2 & 0 \\ 1 & 2 & 0 & -2 \\ 
0 & 0 & -1 & 2
\end{pmatrix}\quad \begin{matrix}
{{[x_{1}, x_{1}]} = 0} \\
{{[x_{1}, x_{4}]} = 0} \\
{{[x_{2}, x_{2}]} = 0} \\
{{[x_{2}, x_{4}]} = 0} \\
{{[x_{3}, x_{3}]} = 0} \\
{{[x_{2}, [x_{1}, x_{3}]]} = {{-\frac{1}{ 3}[x_{3}, [x_{1}, x_{2}]]}}} \\
{{[x_{4}, [x_{3}, x_{4}]]} = 0} \\
{{[[x_{2}, x_{3}], [x_{3}, x_{4}]]} = 0} \\
{{[[x_{1}, x_{3}], [[x_{1}, x_{3}], [x_{3}, x_{4}]]]} = 0}
\end{matrix}
$$
$$
\begin{pmatrix}
2 & -1 & 0 & 0 \\ -1 & 2 & -1 & 0 \\ 0 & -2 & 0 & 3 \\ 
0 & 0 & -1 & 2
\end{pmatrix}\quad \begin{matrix}
{{[x_{1}, x_{3}]} = 0} \\
{{[x_{1}, x_{4}]} = 0} \\
{{[x_{2}, x_{4}]} = 0} \\
{{[x_{3}, x_{3}]} = 0} \\
{{[x_{1}, [x_{1}, x_{2}]]} = 0} \\
{{[x_{2}, [x_{1}, x_{2}]]} = 0} \\
{{[x_{2}, [x_{2}, x_{3}]]} = 0} \\
{{[x_{4}, [x_{3}, x_{4}]]} = 0} \\
{{[[[x_{2}, x_{3}], [x_{3}, x_{4}]], 
 [[x_{3}, [x_{1}, x_{2}]], [[x_{2}, x_{3}], [x_{3}, x_{4}]]]]} = 0}
\end{matrix}
$$
$$
\begin{pmatrix}
2 & -1 & 0 & 0 \\ -2 & 0 & 2 & -1 \\ 0 & 2 & 0 & -1 \\ 
0 & -1 & -1 & 2
\end{pmatrix}\quad \begin{matrix}
{{[x_{1}, x_{3}]} = 0} \\
{{[x_{1}, x_{4}]} = 0} \\
{{[x_{2}, x_{2}]} = 0} \\
{{[x_{3}, x_{3}]} = 0} \\
{{[x_{1}, [x_{1}, x_{2}]]} = 0} \\
{{[x_{3}, [x_{2}, x_{4}]]} = {{-\frac{1}{ 2}[x_{4}, [x_{2}, x_{3}]]}}} \\
{{[x_{4}, [x_{2}, x_{4}]]} = 0} \\
{{[x_{4}, [x_{3}, x_{4}]]} = 0} \\
{{[[x_{1}, x_{2}], [x_{2}, x_{3}]]} = 0}
\end{matrix}
$$
$$
\begin{pmatrix}
0 & 1 & 0 & 0 \\ -1 & 0 & 2 & 0 \\ 0 & -1 & 2 & -1 \\ 
0 & 0 & -1 & 2
\end{pmatrix}\quad \begin{matrix}
{{[x_{1}, x_{1}]} = 0} \\
{{[x_{1}, x_{3}]} = 0} \\
{{[x_{1}, x_{4}]} = 0} \\
{{[x_{2}, x_{2}]} = 0} \\
{{[x_{2}, x_{4}]} = 0} \\
{{[x_{3}, [x_{2}, x_{3}]]} = 0} \\
{{[x_{3}, [x_{3}, x_{4}]]} = 0} \\
{{[x_{4}, [x_{3}, x_{4}]]} = 0} \\
{{[[x_{1}, x_{2}], [[x_{1}, x_{2}], [x_{2}, x_{3}]]]} = 0}
\end{matrix}
$$
$$
\begin{pmatrix}
2 & -1 & 0 & 0 \\ -1 & 2 & -1 & 0 \\ 0 & -2 & 2 & -1 \\ 0 & 0 & -1 & 0
\end{pmatrix}\quad \begin{matrix}
{}[x_{4},  x_{4}] = 0 \\
{}[x_{1},  x_{3}] = 0 \\
{}[x_{1},  x_{4}] = 0 \\
{}[x_{2},  x_{4}] = 0 \\
{}[x_{1},  [x_{1},  x_{2}]] = 0 \\
{}[x_{2},  [x_{1},  x_{2}]] = 0 \\
{}[x_{2},  [x_{2},  x_{3}]] = 0 \\
{}[x_{3},  [x_{3},  x_{4}]] = 0 \\
{}[x_{3},  [x_{3},  [x_{2},  x_{3}]]] = 0
\end{matrix}
$$

$----------------------------------------$

\noindent $\fag_2^{(1)}$ (relations computed up to degree $12$)

$$
\begin{pmatrix}
2 & -1 & 0 & 0 \\ 4 & 0 & -1 & 0 \\ 0 & -1 & 2 & -3 \\ 
0 & 0 & -1 & 2
\end{pmatrix}\quad \begin{matrix}
{{[x_{1}, x_{3}]} = 0} \\
{{[x_{1}, x_{4}]} = 0} \\
{{[x_{2}, x_{2}]} = 0} \\
{{[x_{2}, x_{4}]} = 0} \\
{{[x_{1}, [x_{1}, x_{2}]]} = 0} \\
{{[x_{3}, [x_{2}, x_{3}]]} = 0} \\
{{[x_{4}, [x_{3}, x_{4}]]} = 0} \\
{{[x_{3}, [x_{3}, [x_{3}, [x_{3}, x_{4}]]]]} = 0} \\
{{[[x_{3}, x_{4}], [[x_{1}, x_{2}], [x_{2}, x_{3}]]]} = 
{3\, [[x_{3}, [x_{1}, x_{2}]], [x_{4}, [x_{2}, x_{3}]]]}}
\end{matrix}
$$
$$
\begin{pmatrix}
0 & -4 & 3 & 0 \\ -4 & 0 & 1 & 0 \\ 3 & 1 & 0 & -3 \\ 
0 & 0 & -1 & 2
\end{pmatrix}\quad \begin{matrix}
{{[x_{1}, x_{1}]} = 0} \\
{{[x_{1}, x_{4}]} = 0} \\
{{[x_{2}, x_{2}]} = 0} \\
{{[x_{2}, x_{4}]} = 0} \\
{{[x_{3}, x_{3}]} = 0} \\
{{[x_{2}, [x_{1}, x_{3}]]} = {{-\frac{3}{ 4}[x_{3}, [x_{1}, x_{2}]]}}} \\
{{[x_{4}, [x_{3}, x_{4}]]} = 0} \\
{{[[x_{1}, x_{3}], [x_{3}, x_{4}]]} = 0} \\
{{[[x_{2}, x_{3}], [[x_{2}, x_{3}], [[x_{2}, x_{3}], [x_{3}, x_{4}]]]]} = 0}
\end{matrix}
$$
$$
\begin{pmatrix}
2 & -1 & 0 & 0 \\ -1 & 2 & -1 & 0 \\ 0 & -3 & 0 & 2 \\ 
0 & 0 & -1 & 1
\end{pmatrix}\quad \begin{matrix}
{{[x_{1}, x_{3}]} = 0} \\
{{[x_{1}, x_{4}]} = 0} \\
{{[x_{2}, x_{4}]} = 0} \\
{{[x_{3}, x_{3}]} = 0} \\
{{[x_{1}, [x_{1}, x_{2}]]} = 0} \\
{{[x_{2}, [x_{1}, x_{2}]]} = 0} \\
{{[x_{2}, [x_{2}, x_{3}]]} = 0} \\
{{[[x_{3}, x_{4}], [x_{4}, x_{4}]]} = 0} \\
{{[x_{4}, [[x_{2}, x_{3}], [x_{3}, x_{4}]]]} = 
{{-\frac{1}{ 2}[[x_{3}, x_{4}], [x_{4}, [x_{2}, x_{3}]]]}}}
\end{matrix}
$$
$$
\begin{pmatrix}
2 & -1 & 0 & 0 \\ -3 & 0 & 3 & -1 \\ 0 & 3 & 0 & -2 \\ 
 0 & -1 & -2 & 2
\end{pmatrix}\quad \begin{matrix}
{{[x_{1}, x_{3}]} = 0} \\
{{[x_{1}, x_{4}]} = 0} \\
{{[x_{2}, x_{2}]} = 0} \\
{{[x_{3}, x_{3}]} = 0} \\
{{[x_{1}, [x_{1}, x_{2}]]} = 0} \\
{{[x_{3}, [x_{2}, x_{4}]]} = {{-\frac{1}{ 3}[x_{4}, [x_{2}, x_{3}]]}}} \\
{{[x_{4}, [x_{2}, x_{4}]]} = 0} \\
{{[x_{4}, [x_{4}, [x_{3}, x_{4}]]]} = 0} \\
{{[[x_{1}, x_{2}], [x_{2}, x_{3}]]} = 0}
\end{matrix}
$$
$$
\begin{pmatrix}
1 & -1 & 0 & 0 \\ -2 & 0 & 3 & 0 \\ 0 & -1 & 2 & -1 \\ 0 & 0 & -1 & 2
\end{pmatrix}\quad \begin{matrix}
{}[x_{2},  x_{2}] = 0 \\
{}[x_{1},  x_{3}] = 0 \\
{}[x_{1},  x_{4}] = 0 \\
{}[x_{2},  x_{4}] = 0 \\
{}[x_{3},  [x_{2},  x_{3}]] = 0 \\
{}[x_{3},  [x_{3},  x_{4}]] = 0 \\
{}[x_{4},  [x_{3},  x_{4}]] = 0 \\
{}[[x_{1},  x_{1}],  [x_{1},  x_{2}]] = 0 \\
{}[[x_{1},  x_{2}],  [x_{3},  [x_{1},  x_{2}]]] = -\,  [[x_{2},  x_{3}],  [x_{2},  [x_{1},  x_{1}]]]
\end{matrix}
$$
$$
\begin{pmatrix}
1 & -2 & 0 & 0 \\ -1 & 0 & -1 & 0 \\ 0 & 3 & 2 & -1 \\ 0 & 0 & -1 & 2
\end{pmatrix}\quad \begin{matrix}
{}[x_{2},  x_{2}] = 0 \\
{}[x_{1},  x_{3}] = 0 \\
{}[x_{1},  x_{4}] = 0 \\
{}[x_{2},  x_{4}] = 0 \\
{}[x_{3},  [x_{2},  x_{3}]] = 0 \\
{}[x_{3},  [x_{3},  x_{4}]] = 0 \\
{}[x_{4},  [x_{3},  x_{4}]] = 0 \\
{}[[x_{1},  x_{1}],  [x_{1},  x_{2}]] = 0 \\
{}[[x_{1},  x_{2}],  [x_{3},  [x_{1},  x_{2}]]] = -\,  [[x_{2},  x_{3}],  [x_{2},  [x_{1},  x_{1}]]]
\end{matrix}
$$

$----------------------------------------$

\noindent $\fosp(4|2)^{(2)}$ (relations computed up to degree $12$)

$$
\begin{pmatrix}
1 & -1 & 0 \\ -1 & 0 & 1 \\ 0 & -2 & 2
\end{pmatrix}\quad \begin{matrix}
{{[x_{1}, x_{3}]} = 0} \\
{{[x_{2}, x_{2}]} = 0} \\
{{[x_{3}, [x_{3}, [x_{2}, x_{3}]]]} = 0} \\
{{[[x_{1}, x_{1}], [x_{1}, x_{2}]]} = 0} \\
{{[[x_{1}, x_{2}], [x_{2}, x_{3}]]} = 0}
\end{matrix}
$$

$----------------------------------------$

\noindent $\fp\fsl(3|3)^{(4)}$ (relations computed up to degree $12$)

$$
\begin{pmatrix}
2 & -2 & 0 \\ -1 & 0 & 1 \\ 0 & -2 & 2
\end{pmatrix}\quad \begin{matrix}
{{[x_{1}, x_{3}]} = 0} \\
{{[x_{2}, x_{2}]} = 0} \\
{{[x_{1}, [x_{1}, [x_{1}, x_{2}]]]} = 0} \\
{{[x_{3}, [x_{3}, [x_{2}, x_{3}]]]} = 0} \\
{{[[x_{1}, x_{2}], [x_{2}, x_{3}]]} = 0} \\
{{[[x_{1}, [x_{1}, x_{2}]], [x_{3}, [x_{2}, x_{3}]]]} = 
{{\frac{1}{ 2}[[x_{3}, [x_{1}, x_{2}]], [x_{3}, [x_{1}, x_{2}]]]}}}
\end{matrix}
$$
$$
\begin{pmatrix}
1 & -1 & 0 \\ -1 & 0 & 1 \\ 0 & -1 & 1
\end{pmatrix}\quad \begin{matrix}
{{[x_{1}, x_{3}]} = 0} \\
{{[x_{2}, x_{2}]} = 0} \\
{{[[x_{1}, x_{1}], [x_{1}, x_{2}]]} = 0} \\
{{[[x_{1}, x_{2}], [x_{2}, x_{3}]]} = 0} \\
{{[[x_{2}, x_{3}], [x_{3}, x_{3}]]} = 0} \\
{{[[x_{2}, [x_{1}, x_{1}]], [x_{3}, [x_{2}, x_{3}]]]} = 
{-[[x_{3}, [x_{1}, x_{2}]], [x_{3}, [x_{1}, x_{2}]]]}}
\end{matrix}
$$

$----------------------------------------$

\noindent $\fosp_\alpha(4|2)^{(1)}$, (computed up to degree $12$)

$$
\begin{pmatrix}
2 & 0 & 0 & -1 \\ 0 & 2 & 0 & -\alpha \\ 0 & 0 & 2 & 1+\alpha \\ 
-1 & -1 & -1 & 0
\end{pmatrix}\quad \begin{matrix}
{}[x_{4},  x_{4}] = 0 \\
{}[x_{1},  x_{2}] = 0 \\
{}[x_{1},  x_{3}] = 0 \\
{}[x_{2},  x_{3}] = 0 \\
{}[x_{1},  [x_{1},  x_{4}]] = 0 \\
{}[x_{2},  [x_{2},  x_{4}]] = 0 \\
{}[x_{3},  [x_{3},  x_{4}]] = 0 \\
{}[[x_{2},  x_{4}],  [[x_{1},  x_{4}],  [x_{3},  x_{4}]]] = - 
\frac{\alpha}{1+\alpha}\,  
   [[x_{3},  x_{4}],  [[x_{1},  x_{4}],  [x_{2},  x_{4}]]]
\end{matrix}
$$
$$
\begin{pmatrix}
0 & -1 & -\alpha & 1+\alpha \\ -1 & 0 & 1+\alpha & -\alpha \\ 
-\alpha & 1+\alpha & 0 & -1 \\ 1+\alpha & -\alpha & -1 & 0
\end{pmatrix}\quad \begin{matrix}
{}[x_{1},  x_{1}] = 0 \\
{}[x_{2},  x_{2}] = 0 \\
{}[x_{3},  x_{3}] = 0 \\
{}[x_{4},  x_{4}] = 0 \\
{}[x_{2},  [x_{1},  x_{3}]] = \alpha\,   [x_{3},  [x_{1},  x_{2}]] \\
{}[x_{2},  [x_{1},  x_{4}]] = -(1+\alpha)\,   [x_{4},  [x_{1},  x_{2}]] \\
{}[x_{3},  [x_{1},  x_{4}]] = -\frac{1+\alpha}{\alpha}\,   
{}[x_{4},  [x_{1},  x_{3}]] \\
{}[x_{3},  [x_{2},  x_{4}]] = -\frac{\alpha}{1+\alpha}\,   
{}[x_{4},  [x_{2},  x_{3}]]
\end{matrix}
$$

$----------------------------------------$

\noindent $\fa\fb_3^{(1)}$   (computed up to degree $12$)

$$
\begin{pmatrix}
2 & -3 & 0 & 0 & 0 \\ -1 & 0 & -1 & 0 & 0 \\ 0 & 1 & 2 & -1 & 0 \\ 
0 & 0 & -2 & 2 & -1 \\ 0 & 0 & 0 & -1 & 2
\end{pmatrix}\quad \begin{matrix}
{}[x_{2},  x_{2}] = 0 \\
{}[x_{1},  x_{3}] = 0 \\
{}[x_{1},  x_{4}] = 0 \\
{}[x_{1},  x_{5}] = 0 \\
{}[x_{2},  x_{4}] = 0 \\
{}[x_{2},  x_{5}] = 0 \\
{}[x_{3},  x_{5}] = 0 \\
{}[x_{1},  [x_{1},  x_{2}]] = 0 \\
{}[x_{3},  [x_{2},  x_{3}]] = 0 \\
{}[x_{4},  [x_{3},  x_{4}]] = 0 \\
{}[x_{4},  [x_{4},  x_{5}]] = 0 \\
{}[x_{5},  [x_{4},  x_{5}]] = 0 \\
{}[x_{3},  [x_{3},  [x_{3},  x_{4}]]] = 0 \\
{}[[x_{3},  x_{4}],  [[x_{1},  x_{2}],  [x_{2},  x_{3}]]] = 
    2\,   [[x_{3},  [x_{1},  x_{2}]],  [x_{4},  [x_{2},  x_{3}]]]
\end{matrix}
$$
$$
\begin{pmatrix}
2 & -2 & 0 & 0 & 0 \\ -1 & 0 & 2 & -1 & 0 \\ 0 & 2 & 0 & -1 & -1 \\ 
0 & -1 & -1 & 2 & 0 \\ 0 & 0 & -2 & 0 & 2
\end{pmatrix}\quad \begin{matrix}
{}[x_{2},  x_{2}] = 0 \\
{}[x_{3},  x_{3}] = 0 \\
{}[x_{1},  x_{3}] = 0 \\
{}[x_{1},  x_{4}] = 0 \\
{}[x_{1},  x_{5}] = 0 \\
{}[x_{2},  x_{5}] = 0 \\
{}[x_{4},  x_{5}] = 0 \\
{}[x_{1},  [x_{1},  x_{2}]] = 0 \\
{}[x_{3},  [x_{2},  x_{4}]] = -\frac12\,   [x_{4},  [x_{2},  x_{3}]] \\
{}[x_{4},  [x_{2},  x_{4}]] = 0 \\
{}[x_{4},  [x_{3},  x_{4}]] = 0 \\
{}[x_{5},  [x_{3},  x_{5}]] = 0 \\
{}[[x_{1},  x_{2}],  [x_{2},  x_{3}]] = 0 \\
{}[[x_{2},  x_{3}],  [x_{3},  x_{5}]] = 0
\end{matrix}
$$
$$
\begin{pmatrix}
0 & -3 & 1 & 0 & 0 \\ -3 & 0 & 2 & 0 & 0 \\ 1 & 2 & 0 & -1 & 0 \\ 
0 & 0 & -2 & 2 & -1 \\ 0 & 0 & 0 & -1 & 2
\end{pmatrix}\quad \begin{matrix}
{}[x_{1},  x_{1}] = 0 \\
{}[x_{2},  x_{2}] = 0 \\
{}[x_{3},  x_{3}] = 0 \\
{}[x_{1},  x_{4}] = 0 \\
{}[x_{1},  x_{5}] = 0 \\
{}[x_{2},  x_{4}] = 0 \\
{}[x_{2},  x_{5}] = 0 \\
{}[x_{3},  x_{5}] = 0 \\
{}[x_{2},  [x_{1},  x_{3}]] = -\frac13\,   [x_{3},  [x_{1},  x_{2}]] \\
{}[x_{4},  [x_{3},  x_{4}]] = 0 \\
{}[x_{4},  [x_{4},  x_{5}]] = 0 \\
{}[x_{5},  [x_{4},  x_{5}]] = 0 \\
{}[[x_{2},  x_{3}],  [x_{3},  x_{4}]] = 0 \\
{}[[x_{1},  x_{3}],  [[x_{1},  x_{3}],  [x_{3},  x_{4}]]] = 0
\end{matrix}
$$
$$
\begin{pmatrix}
2 & -3 & 0 & 0 & 0 \\ -1 & 0 & -1 & 0 & 0 \\ 0 & 2 & 2 & -1 & 0 \\ 
0 & 0 & -1 & 2 & -1 \\ 0 & 0 & 0 & -1 & 2
\end{pmatrix}\quad \begin{matrix}
{}[x_{2},  x_{2}] = 0 \\
{}[x_{1},  x_{3}] = 0 \\
{}[x_{1},  x_{4}] = 0 \\
{}[x_{1},  x_{5}] = 0 \\
{}[x_{2},  x_{4}] = 0 \\
{}[x_{2},  x_{5}] = 0 \\
{}[x_{3},  x_{5}] = 0 \\
{}[x_{1},  [x_{1},  x_{2}]] = 0 \\
{}[x_{3},  [x_{2},  x_{3}]] = 0 \\
{}[x_{3},  [x_{3},  x_{4}]] = 0 \\
{}[x_{4},  [x_{3},  x_{4}]] = 0 \\
{}[x_{4},  [x_{4},  x_{5}]] = 0 \\
{}[x_{5},  [x_{4},  x_{5}]] = 0 \\
{}[[[x_{1},  x_{2}],  [x_{2},  x_{3}]],  [[x_{4},  [x_{2},  x_{3}]],  
{}[[x_{1},  x_{2}],  [x_{2},  x_{3}]]]] = 0
\end{matrix}
$$

$----------------------------------------$

\noindent $\fsl(2|4)^{(2)}$, (computed up to degree $21$)

$$
\begin{pmatrix}
2 & -1 & -1 & 0 \\ -1 & 0 & 2 & -1 \\ -1 & 2 & 0 & -1 \\ 0 & -1 & -1 & 2
\end{pmatrix}\quad \begin{matrix}
 {{[x_{2}, x_{2}]}= 0} \\
{{[x_{3}, x_{3}]}= 0} \\
{{[x_{1}, x_{4}]}= 0} \\
  {{[x_{1}, [x_{1}, x_{2}]]}= 0} \\
{{[x_{1}, [x_{1}, x_{3}]]}= 0} \\
  {{[x_{2}, [x_{1}, x_{3}]]}= {[x_{3}, [x_{1}, x_{2}]]}} \\
  {{[x_{3}, [x_{2}, x_{4}]]}= {-\frac{1}{2}\, [x_{4}, [x_{2}, x_{3}]]}} \\
{{[x_{4}, [x_{2}, x_{4}]]}= 0} \\
  {{[x_{4}, [x_{3}, x_{4}]]}= 0}
\end{matrix}
$$
$$
\begin{pmatrix}
2 & -1 & 0 & 0 \\ -2 & 0 & 1 & 1 \\ 0 & -1 & 0 & 2 \\ 0 & -1 & 2 & 0
\end{pmatrix}\quad \begin{matrix}
 {{[x_{2}, x_{2}]}= 0} \\
{{[x_{3}, x_{3}]}= 0} \\
{{[x_{4}, x_{4}]}= 0} \\
  {{[x_{1}, x_{3}]}= 0} \\
{{[x_{1}, x_{4}]}= 0} \\
{{[x_{1}, [x_{1}, x_{2}]]}= 0} \\
  {{[x_{3}, [x_{2}, x_{4}]]}= {[x_{4}, [x_{2}, x_{3}]]}} \\
  {{[[x_{2}, x_{3}], [[x_{1}, x_{2}], [x_{2}, x_{3}]]]}= 0} \\
  {{[[x_{2}, x_{4}], [[x_{1}, x_{2}], [x_{2}, x_{4}]]]}= 0}
\end{matrix}
$$

$$
\begin{pmatrix}
2 & -1 & 0 & 0 \\ -2 & 2 & -1 & 0 \\ 0 & -1 & 0 & 2 \\ 0 & 0 & -1 & 2
\end{pmatrix}\quad \begin{matrix}
 {{[x_{3}, x_{3}]}= 0} \\
{{[x_{1}, x_{3}]}= 0} \\
{{[x_{1}, x_{4}]}= 0} \\
  {{[x_{2}, x_{4}]}= 0} \\
{{[x_{1}, [x_{1}, x_{2}]]}= 0} \\
{{[x_{2}, [x_{2}, x_{3}]]}= 0} \\
  {{[x_{4}, [x_{3}, x_{4}]]}= 0} \\
{{[x_{2}, [x_{2}, [x_{1}, x_{2}]]]}= 0} \\
  {[[[x_{2}, x_{3}], [x_{3}, x_{4}]], [[x_{3}, x_{4}], 
  [[x_{1}, x_{2}], [x_{2}, x_{3}]]]]}= \\ \quad
   {-\, [[[x_{2}, x_{3}], [x_{3}, x_{4}]], [[x_{3}, [x_{1}, x_{2}]], [x_{4}, [x_{2}, x_{3}]]]]}
\end{matrix}
$$

\section*{Table 2.2. Relations for nonsymmetrizable Cartan matrices} 

There are two series of such algebras: $\fsvect^L_\alpha (1|2)$ and
$\fpsq(n)^{(2)}$. 

\noindent $\fpsq(3)^{(2)}$ (computed up to degree $21$)

$$
\begin{pmatrix}
2 & -1 & -1 \\ -1 & 0 & 1 \\ -1 & -1 & 2
\end{pmatrix}\quad \begin{matrix}
{{[x_{2}, x_{2}]} = 0} \\
{{[x_{1}, [x_{1}, x_{2}]]} = 0} \\
{{[x_{1}, [x_{1}, x_{3}]]} = 0} \\
{{[x_{3}, [x_{1}, x_{3}]]} = 0} \\
{{[x_{3}, [x_{2}, x_{3}]]} = 0} \\
{[[x_{2}, [x_{1}, x_{3}]], [x_{2}, [x_{1}, x_{3}]]]} = \\
\qquad\hfill {[[x_{2}, [x_{1}, x_{3}]], [x_{3}, [x_{1}, x_{2}]]] -
[[x_{3}, [x_{1}, x_{2}]], [x_{3}, [x_{1}, x_{2}]]]} \\
{[[x_{1}, [[x_{1}, x_{2}], [x_{2}, x_{3}]]], [[x_{2}, [x_{1}, x_{3}]], [x_{3}, [x_{3}, [x_{1},
x_{2}]]]]]} =\\
-6\, [[[x_{1}, x_{3}], [x_{3}, [x_{1}, x_{2}]]], [[x_{3}, [x_{1}, x_{2}]], [[x_{1}, x_{2}], [x_{2},
x_{3}]]]] \\
-2\, [[[x_{2}, x_{3}], [x_{3}, [x_{1}, x_{2}]]], [[x_{3}, [x_{1}, x_{2}]], [x_{1}, [x_{3}, [x_{1},
x_{2}]]]]]
\end{matrix}
$$
$$
\begin{pmatrix}
0 & 1 & -1 \\ -1 & 0 & 1 \\ 1 & -1 & 0
\end{pmatrix}\quad \begin{matrix}
{{[x_{1}, x_{1}]}= 0} \\
{{[x_{2}, x_{2}]}= 0} \\
  {{[x_{3}, x_{3}]}= 0} \\
{{[[x_{1}, x_{2}], [x_{2}, [x_{1}, x_{3}]]]}= 
   {-\frac{1}{2}\, [[x_{1}, x_{2}], [x_{3}, [x_{1}, x_{2}]]]}} \\
  {{[[x_{1}, x_{3}], [x_{2}, [x_{1}, x_{3}]]]}= {-2\, [[x_{1}, x_{3}], [x_{3}, [x_{1}, x_{2}]]]}} \\
  {{[[x_{2}, x_{3}], [x_{2}, [x_{1}, x_{3}]]]}= {[[x_{2}, x_{3}], [x_{3}, [x_{1}, x_{2}]]]}} \\
  {{[[x_{1}, x_{3}], [[x_{1}, x_{2}], [x_{2}, x_{3}]]]}= 
   {\frac{1}{2}\, [[x_{2}, x_{3}], [[x_{1}, x_{2}], [x_{1}, x_{3}]]] +\frac{3}{4}\, [[x_{2}, [x_{1},
x_{3}]], [x_{3}, [x_{1}, x_{2}]]]}} \\
{[[[x_{1}, x_{2}], [x_{3}, [x_{1}, x_{2}]]], [[x_{3}, [x_{1}, x_{2}] , [[x_{1}, x_{3}], [x_{2},
x_{3}]]]]}= \\ -4\, [[[x_{2}, x_{3}], [x_{3}, [x_{1}, x_{2}]]], [[x_{3}, [x_{1}, x_{2}]], [[x_{1},
x_{2}], [x_{1}, x_{3}]]]]\\ 
+ 4\, [[[x_{1}, x_{3}], [x_{3}, [x_{1}, x_{2}]]], [[x_{3}, [x_{1}, x_{2}]], [[x_{1}, x_{2}], [x_{2},
x_{3}]]]] \\
  {[[[[x_{1}, x_{2}], [x_{1}, x_{3}]], [[x_{1}, x_{3}], [x_{2}, x_{3}]]], [[[x_{1}, x_{2}],
[x_{3}, [x_{1}, x_{2}]]], [[x_{2}, x_{3}], [x_{3}, [x_{1}, x_{2}]]]]]}= \\ 
-\frac{3}{4}\, [[[[x_{1},
x_{2}], [x_{2}, x_{3}]], [[x_{1}, x_{3}], [x_{3}, [x_{1}, x_{2}]]]], 
     [[[x_{1}, x_{3}], [x_{2}, x_{3}]], [[x_{1}, x_{2}], [x_{3}, [x_{1}, x_{2}]]]]] \\ 
    -\frac{1}{2}\, [[[[x_{1}, x_{2}], [x_{2}, x_{3}]], [[x_{1}, x_{3}], [x_{2}, x_{3}]]], 
     [[[x_{1}, x_{2}], [x_{3}, [x_{1}, x_{2}]]], [[x_{1}, x_{3}], [x_{3}, [x_{1}, x_{2}]]]]]
\end{matrix}
$$

$----------------------------------------$

\noindent $psq(4)^{(2)}$, (computed up to degree $21$)

$$
\begin{pmatrix}
2 & -1 & 0 & -1 \\ -1 & 2 & -1 & 0 \\ 0 & -1 & 2 & -1 \\ -1 & 0 & 1 & 0
\end{pmatrix}\quad \begin{matrix}
 {{[x_{4}, x_{4}]}= 0} \\
{{[x_{1}, x_{3}]}= 0} \\
{{[x_{2}, x_{4}]}= 0} \\
  {{[x_{1}, [x_{1}, x_{2}]]}= 0} \\
{{[x_{1}, [x_{1}, x_{4}]]}= 0} \\
  {{[x_{2}, [x_{1}, x_{2}]]}= 0} \\
{{[x_{2}, [x_{2}, x_{3}]]}= 0} \\
  {{[x_{3}, [x_{2}, x_{3}]]}= 0} \\
{{[x_{3}, [x_{3}, x_{4}]]}= 0} \\
  {{[[x_{1}, x_{4}], [x_{3}, x_{4}]]}= 0} \\
  {{[[x_{3}, [x_{1}, x_{4}]], [[x_{3}, x_{4}], [x_{4}, [x_{1}, x_{2}]]]]}= 0} \\
  {{[[x_{4}, [x_{1}, x_{2}]], [[x_{3}, x_{4}], [x_{3}, [x_{1}, x_{2}]]]]}= 
   {[[x_{4}, [x_{2}, x_{3}]], [[x_{1}, x_{4}], [x_{3}, [x_{1}, x_{2}]]]]}} \\
  {[[[x_{3}, [x_{1}, x_{2}]], [[x_{1}, x_{4}], [x_{2}, x_{3}]]], 
    [[[x_{1}, x_{4}], [x_{2}, x_{3}]], [[x_{3}, x_{4}], [x_{4}, [x_{1}, x_{2}]]]]]}= \\ 
\frac{1}{2}\, [[[x_{4}, [x_{1}, x_{2}]], [[x_{1}, x_{4}], [x_{2}, x_{3}]]], 
     [[[x_{1}, x_{4}], [x_{2}, x_{3}]], [[x_{3}, x_{4}], [x_{3}, [x_{1}, x_{2}]]]]] \\ 
+ \frac{1}{2}\, [[[x_{4}, [x_{2}, x_{3}]], [[x_{1}, x_{4}], [x_{2}, x_{3}]]], 
[[[x_{1}, x_{4}], [x_{2}, x_{3}]], [[x_{1}, x_{4}], [x_{3}, [x_{1}, x_{2}]]]]]
\end{matrix}
$$
$$
\begin{pmatrix}
0 & 1 & 0 & -1 \\ -1 & 2 & -1 & 0 \\ 0 & -1 & 0 & 1 \\ -1 & 0 & 1 & 0
\end{pmatrix}\quad \begin{matrix}
 {{[x_{1}, x_{1}]}= 0} \\
{{[x_{3}, x_{3}]}= 0} \\
{{[x_{4}, x_{4}]}= 0} \\
  {{[x_{1}, x_{3}]}= 0} \\
{{[x_{2}, x_{4}]}= 0} \\
{{[x_{2}, [x_{1}, x_{2}]]}= 0} \\
  {{[x_{2}, [x_{2}, x_{3}]]}= 0} \\
{{[[x_{1}, x_{2}], [x_{1}, x_{4}]]}= 0} \\
  {{[[x_{1}, x_{4}], [x_{3}, x_{4}]]}= 0} \\
{{[[x_{2}, x_{3}], [x_{3}, x_{4}]]}= 0} \\
  {{[[x_{3}, [x_{1}, x_{2}]], [x_{3}, [x_{1}, x_{4}]]]}= 0} \\
  {[[x_{4}, [x_{1}, x_{2}]], [[x_{3}, x_{4}], [x_{3}, [x_{1}, x_{2}]]]]}= \\ 
{}[[x_{4}, [x_{3}, [x_{1}, x_{2}]]], [[x_{1}, x_{4}], [x_{2}, x_{3}]]] 
- [[x_{4}, [x_{2}, x_{3}]], [[x_{1}, x_{4}], [x_{3}, [x_{1}, x_{2}]]]] \\
- [[x_{4}, [x_{3}, [x_{1}, x_{2}]]], [[x_{1}, x_{2}], [x_{3}, x_{4}]]] \\
  {[[[x_{4}, [x_{1}, x_{2}]], [[x_{1}, x_{4}], [x_{2}, x_{3}]]], 
    [[[x_{1}, x_{4}], [x_{2}, x_{3}]], [[x_{3}, x_{4}] , [x_{3}, [x_{1}, x_{2}]]]]]}= \\ 
   {[[[x_{4}, [x_{2}, x_{3}]], [[x_{1}, x_{4}], [x_{2}, x_{3}]]], 
    [[[x_{1}, x_{4}], [x_{2}, x_{3}]], [[x_{1}, x_{4}], [x_{3}, [x_{1}, x_{2}]]]]]}
\end{matrix}
$$

$----------------------------------------$

\noindent $\fsvect^L_\alpha(1|2)$, (computed up to degree $12$)

$$
\begin{pmatrix}
2 & -1 & -1 \\ 1 - \alpha & 0 & \alpha \\ 1 + \alpha & -\alpha & 0
\end{pmatrix}\quad \begin{matrix}
{{[x_{2}, x_{2}]} = 0} \\
{{[x_{3}, x_{3}]} = 0} \\
{{[x_{1}, [x_{1}, x_{2}]]} = 0} \\
{{[x_{1}, [x_{1}, x_{3}]]} = 0} \\
{{[[x_{2}, x_{3}], [x_{2}, [x_{1}, x_{3}]]]} = 
 \frac{\alpha+1 }{ \alpha-1 } \, [[x_{2}, x_{3}], [x_{3}, [x_{1}, x_{2}]]]}
\end{matrix}
$$

\section*{\S 3. Comments}

\ssec{3.0} The relations of Tables 2.1 and 2.2 are the defining ones in all
the cases except $\fp\fsl(3|3)^{4)}$ and $\fpsq(3)^{(2)}$. In the last two cases there
are {\it infinitely} many relations that kill the cocycles $c_i$, see sec. 0.8. The degrees of
these relations grow with $i$. (Though these relations look awful when expressed in terms of the
Chevalley generators, they are easy to describe in terms of the matrix units, cf. 0.8.)

Conjectorially, these are the only relations additional to the listed ones.

\ssbegin{3.1}{Statement} {\em (On $3\times 3$ matrices)} There is a relation between
$[x_1, [x_2, x_3]]$ and
$[x_2, [x_3, x_1]]$ for any symmetric Cartan matrix $(A_{ij})$ with
$$
A_{12} + A_{13} + A_{23} = 0.
$$
The relation does not depend on the diagonal elements $A_{ii}$. The relation exists
for any parity of generators and is not reducible to the Serre relations if
$A_{12}A_{13}A_{23}\neq 0$. This relation, together with the Jacobi identity
for
$x_1$, $x_2$ and $x_3$, may be written as
$$
\frac{(-1)^{P(x_1)P(x_3)}} {A_{23}} [x_1, [x_2, x_3]] =
\frac{(-1)^{P(x_1)P(x_2)}} {A_{13}} [x_2, [x_3, x_1]] =
\frac{(-1)^{P(x_2)P(x_3)}} {A_{12}} [x_3, [x_1, x_2]].
$$

For a nonsymmetrizable matrix such that $(A_{ij}=0) \iff (A_{ji}=0)$, this relation
is impossible.

If $x_2$ is odd, $A_{13}=A_{22}=A_{31}=0$ and $A_{23} = -p\, A_{21}$, then the relation
$(ad_{[x_1, x_2]})^p([x_2, x_3])=0$ holds.
\end{Statement}

\ssbegin{3.2}{Remark} It seems that if the ratio $A_{31}:A_{23}$ is a negative rational, but neither
integer nor the inverse of an integer, there is one more relation.
\end{Remark}

\ssbegin{3.3}{Statement} {\em (On $4\times 4$ matrices)} For the Cartan matrix (with
anything instead of each $*$)
$$
\pmatrix * & 1 & p & -pq \\ 1 & * & pq & -p \\
 p & pq & * & -1 \\ -pq & -p & -1 & * \endpmatrix
$$
there is a relation between $[[x_1, x_2], [x_3, x_4]]$, 
\ $[[x_1, x_3], [x_2, x_4]]$ \ and \ $[[x_1, x_4], [x_2, x_3]]$, namely:
$$
pq\, [[x_1, x_2], [x_3, x_4]] - 
(-1)^{P(x_2)P(x_3)}\, q\, [[x_1, x_3], [x_2, x_4]]-
(-1)^{(P(x_2)+P(x_3))P(x_4)}\, [[x_1, x_4], [x_2, x_3]]=0.
$$ 
\end{Statement}

\ssec{3.4. Two statements on $n\times n$ matrices ($n\ge 4$)}

Serre relations involve just two generators. We have seen that even for $\fsl(m|n)$ there are
relations involving 5 generators. It seems that ge

1) Let $\deg x_j=1$ for all
$j$; set
$y_j=x_j^-$. In the free Lie algebra generated by the $x_j$ denote by $v_i$ the expressiona of
degree 1 with respect to {\it each} $x_j$. Comparing the number of equations
$[y_k, \sum c_j v_j]=0$ with the number of parameters ($A_{ij}$ and $c_j$), we see that there exist
relations of degree $n$ involving all the $x_1$, \dots, $x_n$ for
$n \le 5$. We cannot say more about the case $n\ge 6$.

2) If $\fg$ has a central extension, so $\fg^{(m)}_{\varphi }$ has infinitely many
central extensions, and, concequently, infinitely many defining relations: each
central element of positive degree has to be equated to zero. Examples of such
relations are the last indicated relations for $\fp\fsl(3|3)^{(4)}$ and
$\fp\fs\fq(3)^{(2)}$.

\end{document}